\documentclass[a4paper, twocolumn]{article}
\usepackage[utf8]{inputenc}

\usepackage[hmarginratio=1:1,top=32mm,columnsep=20pt]{geometry}

\usepackage{amsmath}
\usepackage{amsfonts}
\usepackage{amssymb}
\usepackage{amsthm}
\usepackage{xcolor}
\usepackage{bm}
\usepackage{mathtools}
\usepackage{graphicx}
\usepackage{subcaption}
\usepackage{float}
\usepackage{commath}
\usepackage[noabbrev]{cleveref}
\usepackage{multirow}
\usepackage{multicol}
\usepackage{caption}
\usepackage{algorithm}
\usepackage{algpseudocode}

\usepackage[
	backend=biber,
	style=numeric,
        maxbibnames=99,
        maxcitenames=1,
        uniquelist=false
	]{biblatex}
\renewbibmacro{in:}{%
  \ifentrytype{article}{}{\printtext{\bibstring{in}\intitlepunct}}}
\DeclareNameAlias{author}{last-first}
\theoremstyle{definition}

\theoremstyle{remark}

\newcommand\numberthis{\addtocounter{equation}{1}\tag{\theequation}}

\newcommand{\Z}{\mathbb{Z}}

\usepackage{authblk}
\title{Prediction of H-Bond Rotations from  Protein H-Bond Topology}
\author[1, 2]{Jørgen Ellegaard Andersen}
\author[2]{Yuki Koyanagi}
\author[3, 4]{Jakob Toudahl Nielsen}
\author[5]{Rasmus Villemoes}
\affil[1]{Danish Institute for Advanced Study, University of Southern Denmark}
\affil[2]{Centre for Quantum Mathematics, Department of Mathematics and Computer Science, University of Southern Denmark}
\affil[3]{Interdisciplinary Nanoscience Center (iNANO), Aarhus University}
\affil[4]{Department of Chemistry, Aarhus University}
\affil[5]{Prevas A/S}
\date{}
\begin{document}

\twocolumn[
\begin{@twocolumnfalse}
\maketitle  
\begin{abstract}
  \noindent
H-bonds are known to play an important role in the folding of proteins into three-dimensional structures, which in turn determine their diverse functions. The conformations around H-bonds are important, in that they can be non-local along the backbone and are therefore not captured by the methods such as Ramachandran plots. We study the relationship between the geometry of H-bonds in proteins, expressed as a spatial rotation between the two bonded peptide units, and their topology, expressed as a subgraph of the protein fatgraph. We describe two experiments to predict H-bond rotations from their corresponding subgraphs. The first method is based on sequence alignment between sequences of the signed lengths of H-bonds measured along the backbone. The second method is based on finding an exact match between the descriptions of subgraphs around H-bonds. We find that 88.14\% of the predictions lie inside the ball, centred around the true rotation, occupying just 1\% of the volume of the rotation space SO(3). 
\end{abstract}
\bigskip
\end{@twocolumnfalse}
]

\section{Introduction}
A protein is a biological molecule consisting of a linear polymer of amino acids. Different proteins fold into unique three-dimensional structures, also called native conformations, and it is widely recognised that the function of a protein is highly dependent on the three-dimensional structure of its native folded state. A classical way to describe the backbone conformation of proteins is the Ramachandran plots, which plots the dihedral angles $(\varphi, \psi)$ before and after each $C^\alpha$ atoms in two-dimensional distributions \cite{ramachandran63}. With the increase in the amount of the available protein structural data, the plots have been updated and extended to be used in structural validation \cites{lovell03, read11} and a number of other purposes (see, for example, \cite{carugo13acta} for a review). A natural extension of Ramachandran plots may be to amalgamate more than one pair of conformation angles to describe the backbone conformation at a larger scale. Levitt \cite{levitt76} proposed a descriptor that combines two consecutive pairs of conformation angles, while Carugo and Djinović-Carugo \cite{carugo13amino} used an average over the dihedral angles $\varphi$ and $\psi$ to characterise entire proteins. Other methods that do not use dihedral angles, include ``curvature'' and ``torsion'' of backbones analogous to the notion of curvature and torsion in differential geometry, computed from the coordinates of $C^\alpha$ atoms \cites{rackovsky78, rackovsky84}, the writhe and the average crossing number in knot theory \cites{rogen03, rogen03new},  projection of nearby atoms to a small sphere centred at each $C^\alpha$ atoms \cite{peng14}, and the coordinates of atoms in two consecutive peptide units \cite{pereira16}.

\begin{figure*}[t]
  \centering
  \begin{subfigure}[b]{.4\linewidth}
    \centering
    \includegraphics[scale=.6]{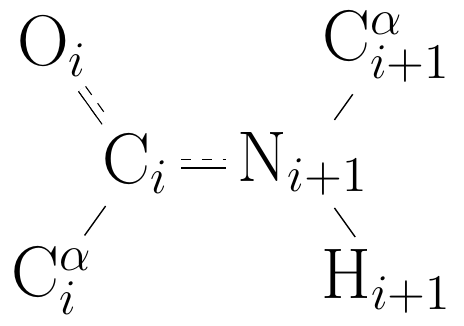}
    \caption{}
    \label{fig:model_pepunit}
  \end{subfigure}
  \begin{subfigure}[b]{.4\linewidth}
    \centering
    \includegraphics[scale=.6]{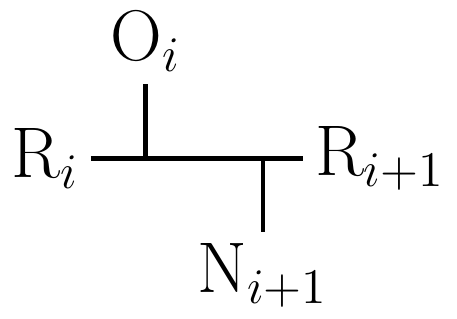}
    \caption{}
    \label{fig:model_block}
  \end{subfigure}
  \par\bigskip
  \begin{subfigure}[b]{.9\linewidth}
    \centering
    \includegraphics[scale=.35]{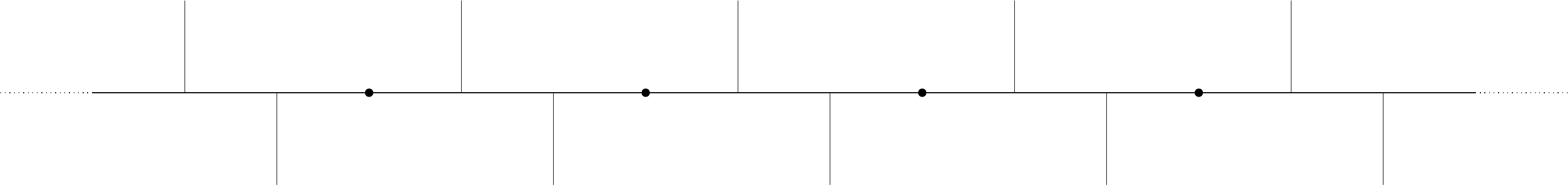}
    \caption{}
    \label{fig:model_backbone}
  \end{subfigure}
  \par\bigskip
  \begin{subfigure}[b]{.9\linewidth}
    \centering
    \includegraphics[scale=.35]{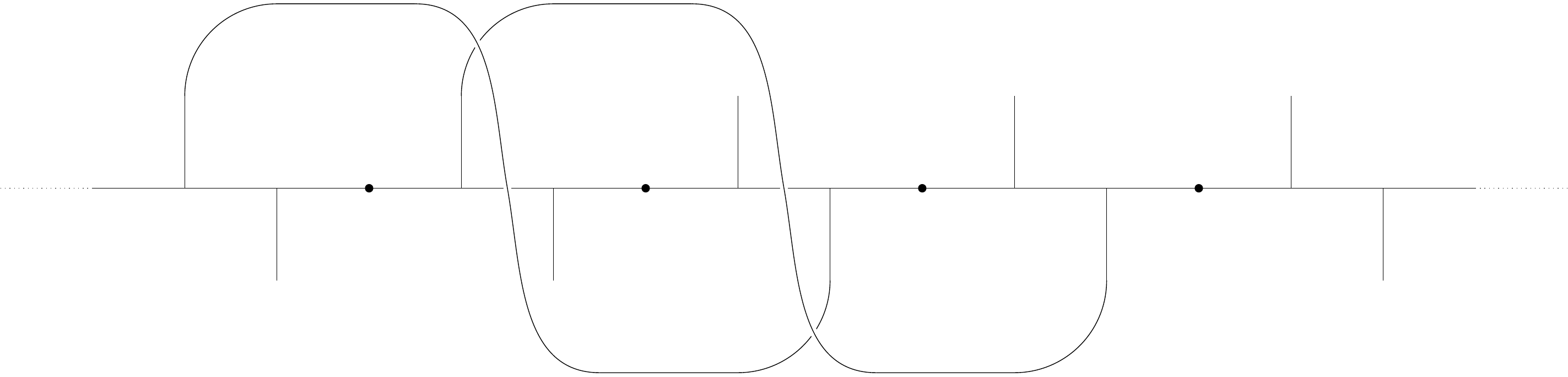}
    \caption{}
    \label{fig:model_prot}
  \end{subfigure}
  \begin{subfigure}[b]{.9\linewidth}
    \centering
    \includegraphics[scale=.35]{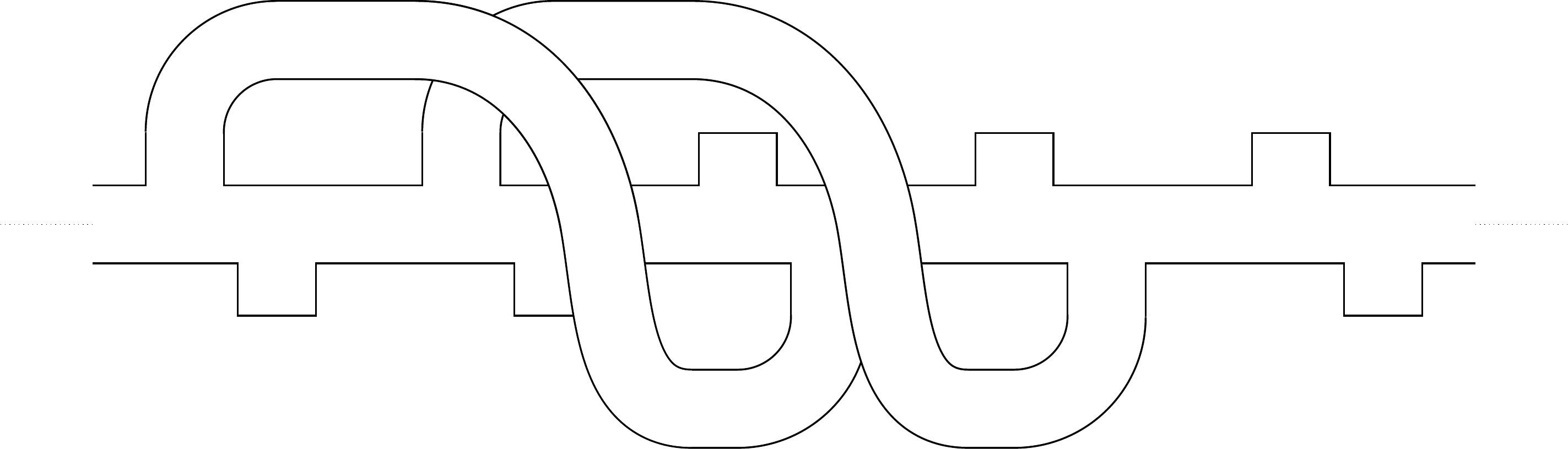}
    \caption{}
    \label{fig:model_surface}
  \end{subfigure}
  \caption{Construction of the protein model. Each peptide unit (\subref{fig:model_pepunit}) is represented by a building block with the donor nitrogen and the acceptor oxygen half-edges drawn below and above the backbone, respectively (\subref{fig:model_block}). These building blocks are concatenated at so-called $\alpha$-carbon linkages to form a model of the backbone (\subref{fig:model_backbone}). The H-bonds are drawn as edges between corresponding donor and acceptor half-edges (\subref{fig:model_prot}). One can associate a surface to a given fatgraph (\subref{fig:model_surface}).}
  \label{fig:model}
\end{figure*}

A common feature for the above methods is that they describe conformation of a backbone by moving along it from one end to the other. One of the main mechanisms determining and stabilising the folded structure is H-bonds \cite{bordo94, rose93, pace14}, whose geometry has also been studied in relation to the structural analysis of proteins \cites{grishaev04, morozov04}. Spatial rotations were introduced as a systematic three-dimensional descriptor of H-bond geometry in \cite{penner14}, and were found to correspond well to the concrete secondary structures and other local structural motifs \cite{penner14}. If we ignore the geometry of H-bonds and instead concentrate on their topology, we obtain a graph, with backbone atoms as vertices and the covalent bonds and H-bonds as edges. In \cite{penner10}, an extension to this structure, called a \emph{fatgraph}, was used to study protein structures. A fatgraph can be thought of as a surface obtained by ``fattening'' the underlying graph; i.e.\ by expanding the vertices to discs, and edges to ``ribbons'' connecting these discs (\Cref{fig:model_surface}). In this paper, we utilise the construction of protein fatgraphs described in \cite{penner10}, with some minor simplifications. The construction is summarised in \Cref{fig:model}. The model is based on a representation of peptide units as fatgraph ``building blocks'' (\Cref{fig:model_block}), which are then glued together to represent a backbone (\Cref{fig:model_backbone}). Finally, the H-bonds are added as edges between the donor and the acceptor half-edges (\Cref{fig:model_prot}).

The correspondence between fatgraphs and surfaces allows us to compute topological invariants, such as genus and the number of boundary components, for protein structures. In \cite{penner11} the relation between these topological invariants and the protein domains was discussed. In addition, the planarity of peptide units allows us to assign an element of the rotation group $\mathrm{SO(3)}$ to each H-bond, as described in \cite{penner14} (See Supplementary Material for a detailed description). In \cite{penner14}, it was discovered that the H-bond rotations computed from PDB \cite{berman00} are concentrated around well-defined clusters, and that these clusters correlated with some well-known local structures.

In this paper, we study the relation between topology and geometry of proteins around each H-bond. The study forms a part of an effort to solve the problem of predicting proteins' geometric structure from their primary structures, also called the protein folding problem. Our intended approach is in two steps, and relies on the fatgraph structure as an intermediate step. In the first stage a fatgraph structure is predicted from a primary sequence, and in the second stage a geometric structure is predicted from a protein fatgraph. Progress is being made on the different aspect of the programme, including the enumeration of possible protein fatgraph structures \cite{andersen21enum}, and the study of relationship between protein fatgraphs and geometric structures \cites{andersen21using, andersen21topo}.

\begin{figure}[t]
  \centering
  \includegraphics[width=\linewidth]{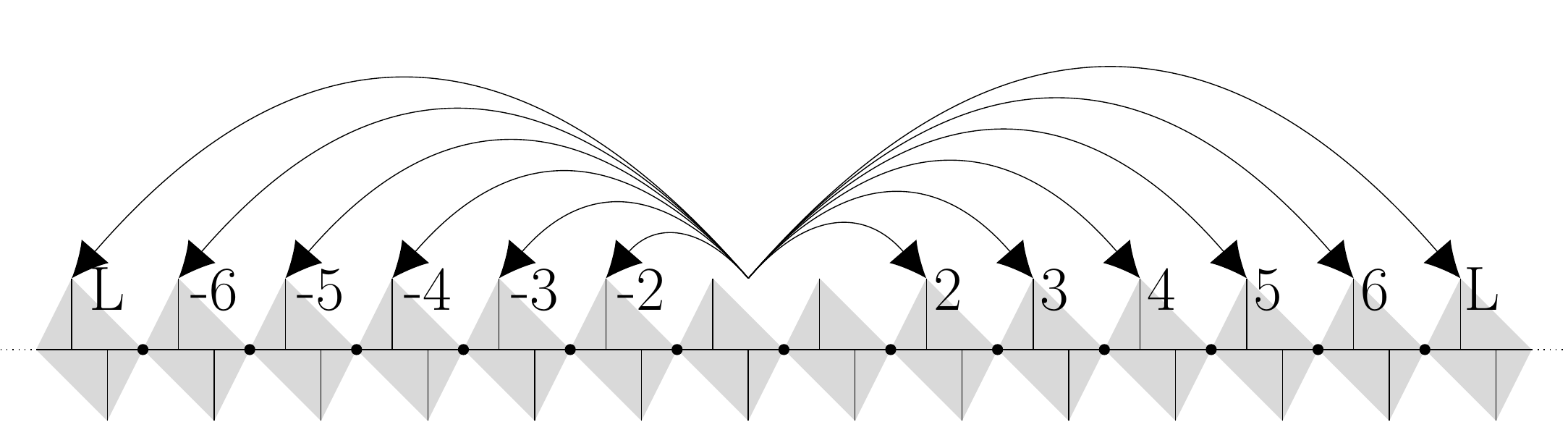}
  \caption{Signed length of H-bonds. Peptide units are shaded grey. All bonds longer than $\pm 6$ are denoted by L.}
  \label{fig:signedlength}
\end{figure}

The current study is a contribution to the second stage, but instead of predicting the entire geometric structure from a protein fatgraph, we aim to predict local geometry of proteins, expressed as spatial rotations along the H-bonds, using fatgraph structures around the H-bonds. We call these local fatgraph structures \emph{H-bond local patterns}, or simply \emph{H-bond patterns}. For a given H-bond $a$, the \emph{H-bond local pattern} or simply the \emph{H-bond pattern} of \emph{window size} $w$ around $a$ is the sub-fatgraph of the protein fatgraph consisting of the set of backbone atoms, whose distance along the backbone to one of the endpoints of $a$ is no more than $w$ atoms, together with all backbone and H-bond edges between them. We call $a$ the \emph{central bond} of the pattern. The central bond determines the \emph{signed length} of the bond, which is the distance between the central bond's endpoints, measured from the donor to the acceptor. Formally, if the donor of the bond is in the $i$'th peptide unit and the acceptor in the $j$'th peptide unit, the the signed distance $d$ is defined to be $j-i$ if $i<j$ and $j-i-1$ otherwise (\Cref{fig:signedlength}).

\begin{figure*}[t]
  \centering
  \begin{subfigure}[c]{1\linewidth}
    \centering
    \includegraphics[scale=.6]{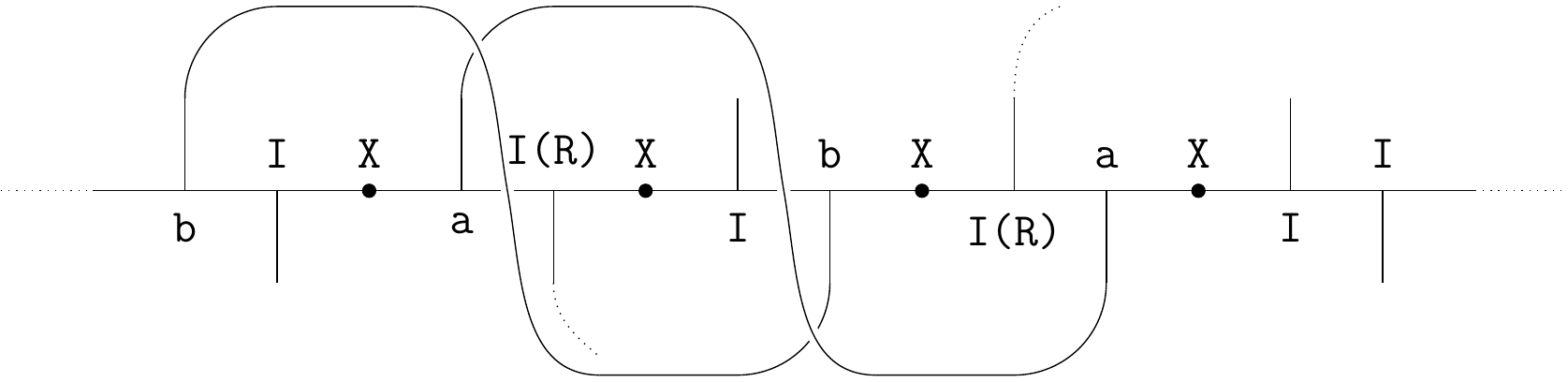}
    \caption{Connected pattern. The pattern can be expressed as \texttt{bIXaIXIbXIaXII\_-3}.}
    \label{fig:locpat_c}
  \end{subfigure}
  \begin{subfigure}[c]{1\linewidth}
    \centering
    \vspace{2em}
    \includegraphics[scale=.45]{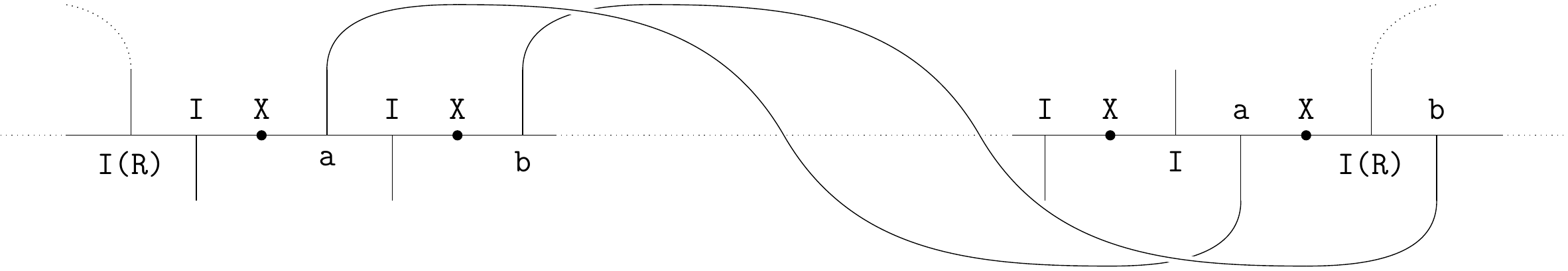}
    \caption{Disconnected pattern. The pattern can be expressed as \texttt{IIXaIXb:IXIaXIb\_L}.}
    \label{fig:locpat_nc}
  \end{subfigure}
  \caption[H-bond patterns]{Connected (\subref{fig:locpat_c}) and disconnected (\subref{fig:locpat_nc}) H-bond local patterns of window size 3. The parts of backbone and H-bonds, that do not participate in the pattern are shown by dotted lines. Note the remotely-bonded atoms may be replaced by the letter \texttt{R}, depending on the parameter specified (see text for details).}
  \label{fig:hb_examples}
\end{figure*}

We will also record the H-bonds whose two endpoints are no more than $w$ atoms away from one of the endpoints of $a$ along the backbone. An H-bond pattern may be expressed as a sequence of letters as follows (\Cref{fig:locpat_c});
\begin{equation*}
\texttt{bIXaIXIbXIaXII}
\end{equation*}
Each pair of lowercase letters indicate a pair of atoms with an H-bond between them. The letter ``\texttt{a}'' is given a special meaning as an indication of the central bonds. The remaining H-bonds are ordered by the position of the donor atoms (starting from the N-terminus) and given letters ``\texttt{b}, \texttt{c}, ... ''. The uppercase letter \texttt{X} indicates a C$^\alpha$ atom, and the uppercase letter \texttt{I} indicates an ``isolated'' N or O half-edge (with no H-bond attached). Hence we see that the above pattern corresponds to the structure shown in \Cref{fig:locpat_c}, with the second H-bond as the central bond. If the distance between the two endpoints of the central bond is greater than $2w$, the H-bond pattern is not connected, which may be indicated by ``\texttt{:}''. So we may obtain a pattern such as (\Cref{fig:locpat_nc});
\begin{equation*}
\texttt{IIXaIXb:IXIaXIb}
\end{equation*}

This local pattern description can be enhanced further by encoding various extra pieces of information:

\begin{itemize}

\item A backbone atom within the window may have an H-bond
attached, whose other endpoint lies outside the window. In
the basic pattern description, this would be indistinguishable from an unbonded backbone atom; using the letter \texttt{R} (for \texttt{R}emotely bonded) instead of \texttt{I} allows us to distinguish these cases. We may also restrict this information for atoms located at most $r$ from either endpoint of the central bond, for some $0\leq r\leq w$. Atoms which are remotely bonded but lie further than $r$ from the central bond are then treated as if they were isolated atoms.

\item We may record whether an H-bond is twisted or not. Formally, this is determined by whether the inner product between the normal vectors to the peptide planes is positive or negative. If the inner product is negative (that is, if the bond is twisted), we replace the lowercase letter with the corresponding letter from the other end of the alphabet (that is, \texttt{z} instead of \texttt{a}, \texttt{y} instead of \texttt{b} etc.). Similarly to the remote bonds, this information may be restricted to the atoms located at most $t$ atoms from either endpoint of the central bond, for some $0 \leq t \leq w$.

\item Beside the local pattern itself, we may record separately the residues at the four C$^\alpha$'s closest to the central bond's endpoints. To reduce the number of possible bond descriptions and obtain reasonable clusters, we choose a grouping of the 20 residues into 1, 2, 3 or 4 groups according to their chemical properties (Supplementary Material), and simply record the 4-tuple of group identifiers.  The four residues around an H-bond are chosen and recorded in the following order;
\begin{enumerate}
\item The residue preceding the N-donor amino acid residue.
\item The N-donor amino acid residue.
\item The O-acceptor amino acid residue.
\item The residue following the O-acceptor amino acid residue.
\end{enumerate}

\end{itemize}

The resulting pattern description may look as follows:
\begin{align*}\label{pattern}
  &\texttt{10XbRXcRXdRXzbXvcXIdXIvXRzXRRXRRXRRX}\\
  &\qquad\texttt{\_5\_4LLLL}\numberthis
\end{align*}
Here the number \texttt{10} indicates the window size, and the central bond is twisted; as shown by the use of the letter \texttt{z} instead of \texttt{a}. There are several remotely-bonded atoms, indicated by the letter \texttt{R}. The last segment, \texttt{4LLLL}, indicates the number of groups used in the grouping of residues and the four group identifiers.

The above description of H-bond patterns only relies on the fatgraph structure of a protein, with its backbone and H-bonds. It is also possible to generate a local pattern which includes information about a protein's tertiary contacts, by adding an edge to the fatgraph, where there is a tertiary contact. These edges may be labelled to indicate their status as tertiary contacts. 

In the context of protein structure research, in particular the protein structure prediction, the effectiveness of deep learning approach has been demonstrated by AlphaFold and AlphaFold2 systems \cites{senior20, jumper20} in CASP \cite{moult95}. The current study differs from the protein structure prediction as studied in CASP in two points. Firstly, our input is the fatgraph structure of proteins, and not their primary sequences. Our method is not dependent on the primary sequence and the information on residues only plays an ancillary role. Secondly, our output is a spatial rotation along a given H-bond, a descriptor of the local structure around the H-bond. Our main focus is to study the relationship between the topology and the geometry of proteins, and to investigate the use of spatial rotation as a descriptor of protein structures. 

\section{Methods and Results}
\subsection{Dataset}

\begin{figure*}[t]
  \centering
  \includegraphics[scale=.6]{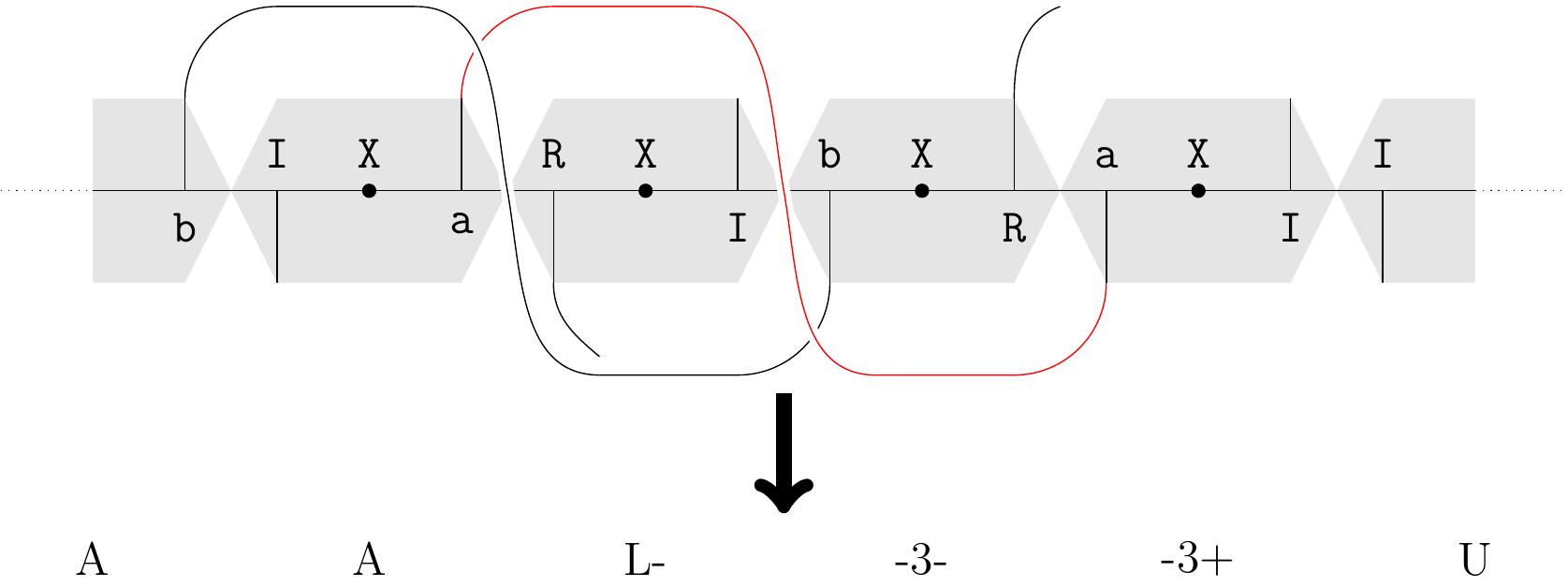}
  \caption[H-bond lengths sequence]{Translation of H-bond pattern to H-bond length sequence. Amino acids  are shaded grey. A twisted bond is shown in red. }
  \label{fig:h_seq}
\end{figure*}

  The dataset used was based on the HQ60 dataset in \cite{penner14}. Here we give a brief definition of this dataset. PISCES \cite{wang03} is a service that, among other things, creates subsets of sequences from PDB based on specified threshold for structure quality and sequence identity. For the HQ60 dataset, we use only X-ray structures, with a resolution threshold of 2.0Å, Rfac threshold of 0.2, and maximum sequence homology of 60\%. The data was extracted from PDB \cite{berman00} in March 2021, resulting in a collection of 16262 proteins. The H-bonds are taken from the DSSP program \cite{kabsch83}, with the additional conditions \cite{baker84};
  \begin{align*}
    &\text{HO-distance} < 2.7\text{Å} \\
    &\text{angle(NHO)}, \text{angle(COH)} > 90^\circ.
  \end{align*}
  The resulting dataset contained approximately 2.4 million H-bonds.

  A subset of 200 proteins was randomly selected as the test data from the dataset, and the remaining 16,062 proteins were used as the training data.
  
  \subsection{H-bond pattern alignment}\label{sec:method_align}

  For each H-bond in the training data, H-bond local pattern was generated using the following parameter combination;
  
\begin{itemize}
\item Window size: 10
\item Remote bonds: 10
\item Twisted bonds: 10
\item Number of residue classes: 1
\end{itemize}

Each H-bond pattern was translated to a sequence, which we call an H-bond length sequence, based on the lengths of the H-bonds. The translation was done as follows (see \Cref{fig:h_seq}).

\begin{enumerate}
\item Recall that each H-bond pattern corresponds to a repeated sequence of N-$\mathrm{C^\alpha}$-O atoms, representing an amino acid, in our protein model. At either end of the pattern we may have only a part of this three-atoms sequence; for example it may start with a $\mathrm{C^\alpha}$ or an O atom. Given an H-bond pattern, we split it into three-atom segments corresponding to amino acids (we may end up with segments at either end of the pattern which consist of fewer than three atoms). In the following procedure we only consider the atoms in the H-bond pattern. 
\item For each of the resulting segments, check whether the N atom is a donor. If so, we assign a symbol encoding the signed length followed by the twistedness of the bond (``+'' for twisted, and ``-'' for not twisted). E.g.\ ``+4-'' for a bond of length +4, which is not twisted (``-'').
\item For each of the remaining segments, check whether the O atom is an acceptor. If so, we assign a symbol ``A'' to the segment.
\item The remaining segments are assigned a symbol ``U'' for unbonded.
\end{enumerate}

We note the H-bond length sequences constructed as above completely determine the local H-bond fatgraph structure of proteins. Using the above translation method, we can compute the list of H-bond length sequence and associated rotation value from our training dataset. The same procedure is applied to the proteins whose H-bond rotations are to be predicted. For an H-bond local pattern of window size 10 around the central bond $a$ in the prediction dataset, let $s(a)$ be the corresponding H-bond length sequence (whose length is the number of whole or partial amino acids contained in the H-bond pattern). We then compute alignment score for $s(a)$ against each of the H-bond length sequences in the training dataset, using Needleman-Wunsch algorithm \cite{needleman70}, with the substitution matrix constructed as follows.
Let
\begin{align*}
  S_1 &= \{-6-, -5-, \dotsc, +5-, +6-, \\
  &\qquad -6+, -5+, \dotsc, +5+, +6+\} \\
  S_2 &= \{L-, L+\} \\
  S &= S_1 \cup S_2 \\
  K &= S \cup \{ \mathrm{U, A}\}.
\end{align*}
Define $l: S \to \Z \cup \{L\}$ to be the function that returns the length part of $x \in S$, i.e.\, it removes the last character in $x$. Define also $t: S \to \{+, -\}$ to be the function that returns the twistedness of $x \in S$. Construct the substitution matrix $M^{(1)}$ with entries $M^{(1)}_{k_1, k_2}$, $k_1, k_2 \in K$ by the pseudocode shown in \Cref{alg:subst_matrix}.
\begin{algorithm}
\begin{algorithmic}[1]
  \For{$k_1 \in K$}
  \For{$k_2 \in K$}
  \If{$k_1==k_2$}
  \State $s=1$
  \ElsIf{$\{k_1,k_2\} \cap (K \setminus S) \neq \emptyset$}
  \State $s=-1$
  \ElsIf{$l(k_1) == l(k_2)$}
  \State $s=0$
  \ElsIf{$\{k_1, k_2\} \cap S_2 \neq \emptyset$}
  \State $s=-0.75$
  \Else{}
  \State $s=-\abs{l(k_1)-l(k_2)}/20$
  \EndIf
  \If{$\{k_1, k_2\} \subset S$ and $t(k_1) \neq t(k_2)$}
  \State $s = s-0.1$
  \EndIf
  \State $M^{(1)}_{k_1,k_2} = s$
  \EndFor
  \EndFor
\end{algorithmic}
\caption{Pseudocode for the construction of substitution matrix $M^{(1)}$.}
\label{alg:subst_matrix}
\end{algorithm}
The gap score was set to -1. 

Our prediction for the rotation along $a$ is the rotation value associated to the H-bond length sequence with the highest alignment score for  $s(a)$.

Three further substitution matrices, $M^{(2)}, M^{(3)},$ and $M^{(4)}$ were constructed to investigate the effect of different penalties for mismatch. In $M^{(2)}$ the penalty for when $k_1, k_2 \in S_1$ (``short to short'' substitution) was made exponential instead of linear, by replacing line 12 with
\begin{algorithmic}
  \State $d=-\abs{l(k_1)-l(k_2)}$
  \State $s = -0.6 \left((\exp{d} - 1) / (\exp{12} - 1) \right)$.
\end{algorithmic}
In $M^{(3)}$ the penalty was made logarithmic by replacing line 12 with
\begin{algorithmic}
  \State $d=-\abs{l(k_1)-l(k_2)}$
  \State $s = -0.6 \left(\log{(d + 1)} / \log{(12 + 1)} \right)$.
\end{algorithmic}
Finally, in $M^{(4)}$, the penalty for different twistedness values was increased by replacing line 15 by
\begin{algorithmic}
  \State $s = s - 0.8$.
\end{algorithmic}

The effect of different gap scores was investigated by using the simplest substitution matrix; 1 along the diagonal (match) and -1 elsewhere (mismatch), with different gap scores. We call these simple substitution matrices $M_{\text{simple}}^{(p)}$, where $p$ is the gap penalty score.

\subsection{H-bond pattern matching}
In the second method, we attempt to find an exact match for a given H-bond local pattern in the training data. To ensure we find a match, we generate a sequence of local patterns ranging in their complexity.

For each H-bond in the training data, H-bond local patterns were constructed with the following parameter combinations;

\begin{itemize}
\item Window size: 0 (only the central bond), $1, 2, \dotsc, 10$
\item Remote bonds: 0 (do not indicate remote bond), $1, 2, \dotsc,$  window size
\item Twisted bonds: none (-1), only the central bond (0), window size
\item Number of residue classes: 1, 2, 3, 4
\end{itemize}

This resulted in 792 parameter combinations. For each of these parameter combinations, the H-bond patterns with less than 30 occurrences were discarded. For each of the remaining H-bond patterns with the associated $\mathrm{SO}(3)$ rotations, we performed a clustering analysis. A description of the clustering algorithm can be found in \cite{penner14} (Method section). We note that rotations are represented as axis-angle pairs, and the algorithm uses a discretised rotational space; it divides the cube $(-\pi, \pi)^3$ into $81 \times 81 \times 81$ small boxes, and finds a mode box for each well-defined cluster. Each box can belong to at most one cluster, even when the algorithm finds several clusters.

We then compute the score $s$ for each clustering run by the following formula;
\begin{equation}
  \label{eq:score}
  s =
  \begin{cases}
    \pi - m & \text{if there is only one cluster} \\
    -1 & \text{if there are } >1 \text{ cluster},
  \end{cases}
\end{equation}
where $m$ is the mean distance between all boxes in the cluster and the mode box, which of course is bounded by $\pi$. In this way we associate a score for each bond description. The result is a table where each row contains an H-bond pattern, a rotation value and a score. This data is then used to predict rotation of a given H-bond from a topological model of the protein, which we describe below.

Prediction was done on the test data of 200 proteins, using the same procedure as the training. For each H-bond in the test data, we obtain H-bond patterns using the same sets of parameter combinations as used in the training stage. Each resulting H-bond pattern is looked up in the table of patterns for which a clustering was performed. If a match is found, we obtain an estimate for the bond's rotation, which is the centre of the largest cluster, along with a score for that estimate, which is the score associated to the cluster. Our final prediction for the rotation associated to the H-bond is the estimate with the highest associated score. If two estimates have the same score, the result with more detailed H-bond pattern is used for the prediction.

In order to promote the selection of larger pattern for prediction, we modified the score function (\ref{eq:score}) to penalise the use of smaller patterns. The new score function is given by

\begin{equation}
  \label{eq:score2}
  s =
  \begin{cases}
    \pi - &m - \exp(3-w) \\
    &\text{if there is only one cluster} \\
    -1 & \text{otherwise},
  \end{cases}
\end{equation}
where $m$ is the mean distance to the cluster mode, and $w=\max\{3, \text{window size}\}$.

The above analysis was repeated  using the local H-bond patterns generated including the tertiary contact information, where the edges corresponding to tertiary contacts and backbone $\alpha$-carbon linkages contribute equally to the computation of window size, and we do not consider atoms more than one tertiary contact away from the central bond. In other words, an atom is only included in the H-bond pattern if the distance (along the backbone and the tertiary contact edges) from it to either end of the central bond is less than or equal to the window size, and that no more than one tertiary contact is traversed in computing the distance. This criteria was applied to limit the pattern to the atoms most likely to exert some influence on the central H-bond.

\subsection{Results}
\begin{figure*}[t]
  \centering
  \begin{subfigure}[c]{.3\linewidth}
    \includegraphics[width=\textwidth]{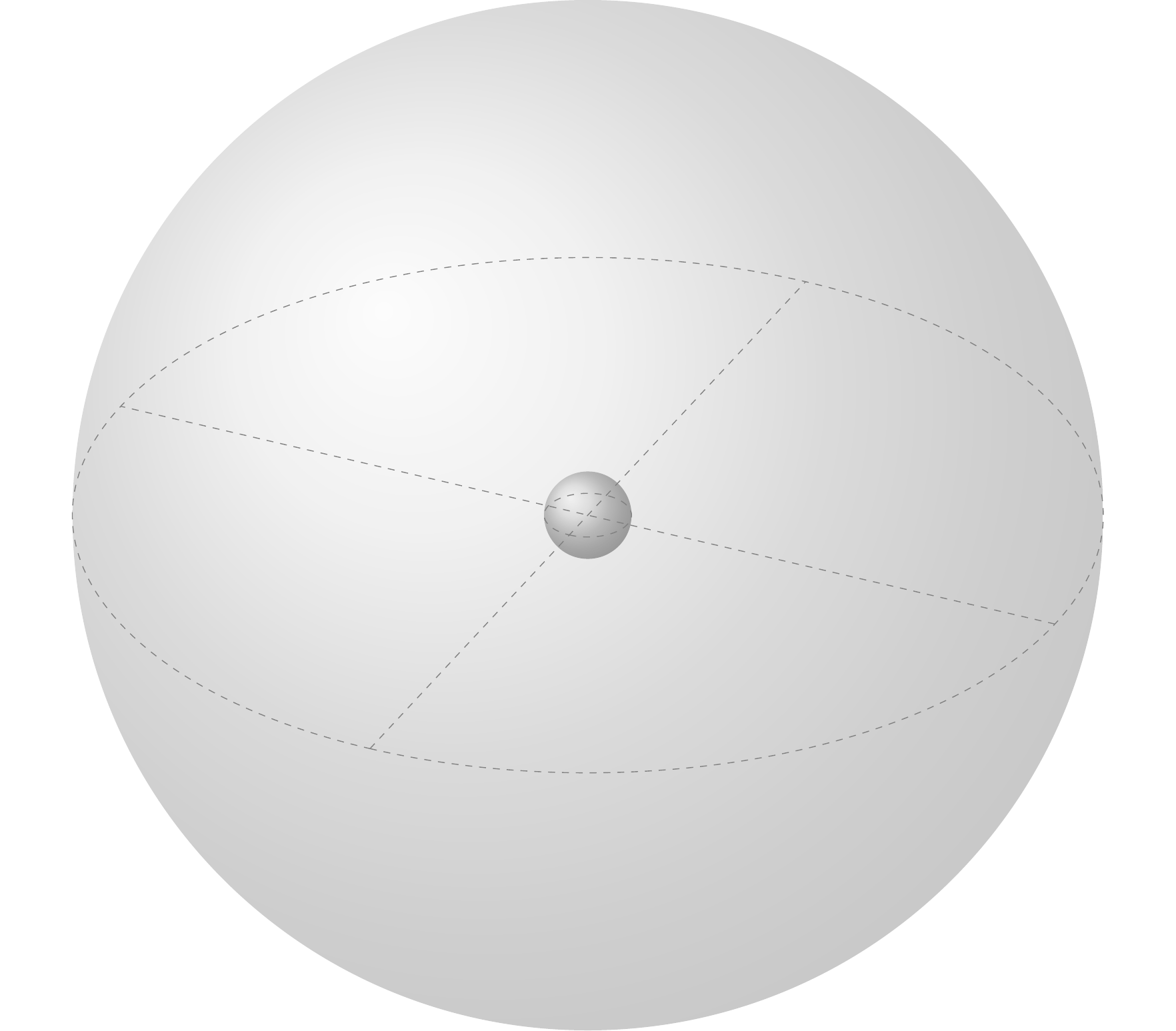}
    \caption{}
    \label{fig:so3_a}
  \end{subfigure}
  \begin{subfigure}[c]{.3\linewidth}
    \includegraphics[width=\textwidth]{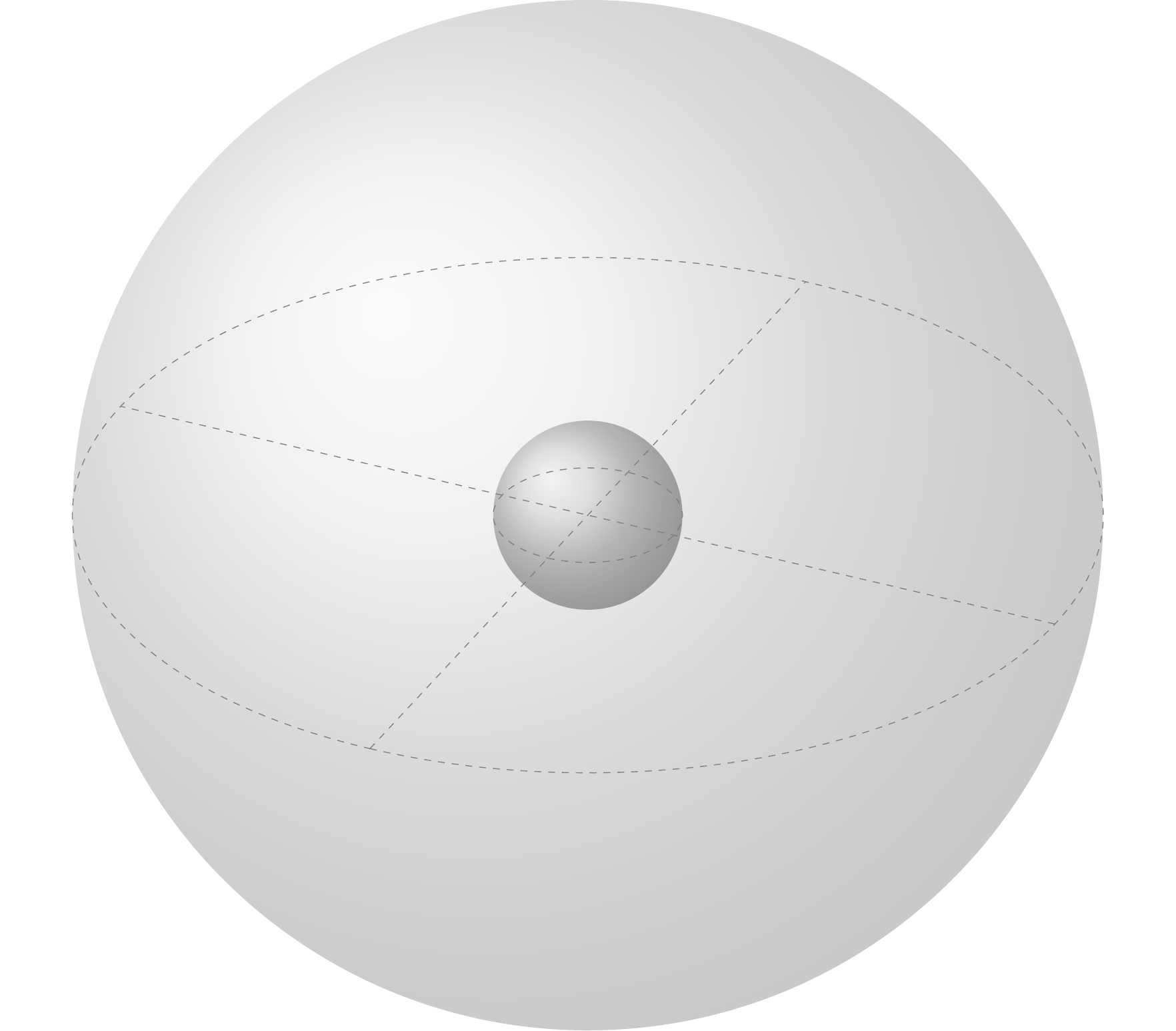}
    \caption{}
    \label{fig:so3_b}
  \end{subfigure}
  \begin{subfigure}[c]{.3\linewidth}
    \includegraphics[width=\textwidth]{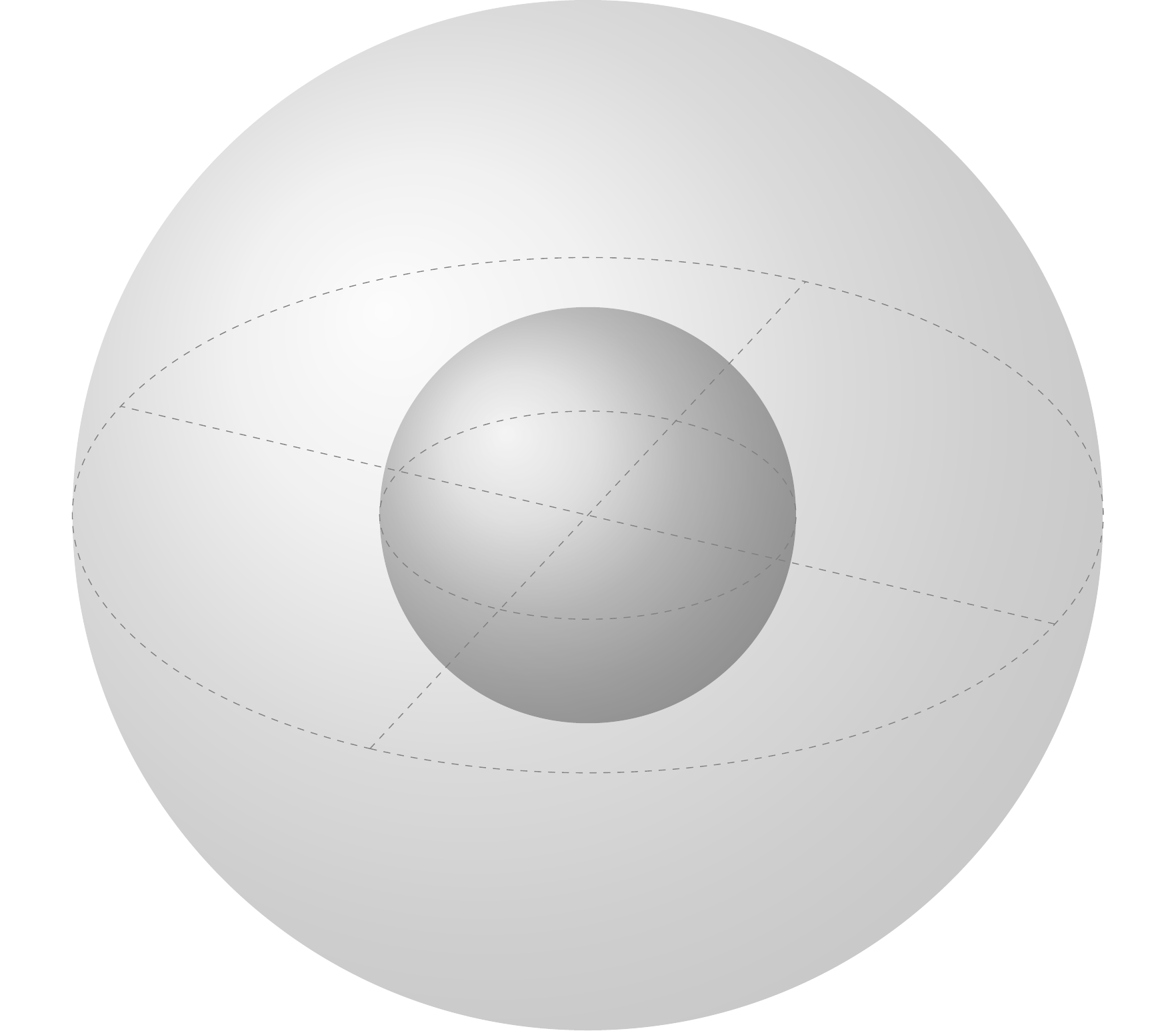}
    \caption{}
    \label{fig:so3_c}
  \end{subfigure}
  \caption[SO(3) volumes]{Illustrations of balls comprising 0.1\% (\subref{fig:so3_a}), 1\% (\subref{fig:so3_b}) and 10\% (\subref{fig:so3_c}) of the total volume of SO(3). The entire SO(3) is shown as a ball of radius $\pi$; a point $p$ in the ball defines a rotation via the angle-axis pair $(\theta, v)$, where $\theta = \norm{p}$ and $v=p/\norm{p}$.}
  \label{fig:so3}
\end{figure*}

\begin{table*}[t]
  \centering
  \begin{tabular}{*{6}{r}}
    \hline
    \multicolumn{2}{c}{Range of $d$} & $M^{(1)}$ & $M^{(2)}$ & $M^{(3)}$ & $M^{(4)}$ \\
    \hline
    $d < 0.2664$ &  (0.1\%) & 49.56 & 52.57 & 52.62 & 52.66 \\
    $d < 0.4567$ &  (0.5\%) & 66.49 & 72.59 & 72.66 & 72.70 \\
    $d < 0.5766$ &  (1.0\%) & 72.91 & 79.63 & 79.71 & 79.77 \\
    $d < 0.7862$ &  (2.5\%) & 80.65 & 87.08 & 87.16 & 87.30 \\
    $d < 0.9968$ &  (5.0\%) & 85.33 & 90.97 & 91.02 & 91.14 \\
    $d < 1.2689$ & (10.0\%) & 89.17 & 93.92 & 93.94 & 94.11 \\
    $d < 1.7663$ & (25.0\%) & 94.49 & 97.07 & 97.04 & 97.24 \\
    $d < 2.3099$ & (50.0\%) & 97.48 & 98.61 & 98.60 & 98.84 \\
    $d < 2.7437$ & (75.0\%) & 98.86 & 99.35 & 99.36 & 99.50 \\
    $d < 3.1416$ & (100\%)  &100.00 &100.00 &100.00 & 100.00\\
    \hline
  \end{tabular}
  \caption{Accumulative \% of H-bonds whose predicted rotation values lies within the specified distance from the true rotation values, for different substitution matrices $M^{(1)}, M^{(2)}, M^{(3)}$ and $M^{(4)}$ (see \cref{sec:method_align} for the details of how these are constructed). The numbers in parentheses show the volume of the ball whose radius is the upper limit of the range as a proportion of the volume of entire SO(3), w.r.t.\ the invariant metric.}
  \label{tab:align_subst}
\end{table*}

\begin{table*}[t]
  \centering
  \begin{tabular}{*{5}{r}}
    \hline
    \multicolumn{2}{c}{Range of $d$} & $M_{\text{simple}}^{(-1)}$  & $M_{\text{simple}}^{(-5)}$ & $M_{\text{simple}}^{(0)}$ \\
    \hline
    $d < 0.2664$ & (0.1\%)  & 52.84 & 53.52 & 50.27 \\
    $d < 0.4567$ & (0.5\%)  & 72.82 & 73.69 & 69.01 \\
    $d < 0.5766$ & (1.0\%)  & 79.94 & 80.97 & 75.92 \\
    $d < 0.7862$ & (2.5\%)  & 87.37 & 88.43 & 83.31 \\
    $d < 0.9968$ & (5.0\%)  & 91.24 & 92.26 & 87.71 \\
    $d < 1.2689$ & (10.0\%) & 94.13 & 95.13 & 91.01 \\
    $d < 1.7663$ & (25.0\%) & 97.15 & 97.64 & 95.17 \\
    $d < 2.3099$ & (50.0\%) & 98.82 & 99.04 & 97.68 \\
    $d < 2.7437$ & (75.0\%) & 99.51 & 99.61 & 99.01 \\
    $d < 3.1416$ & (100\%) & 100.00 & 100.00 & 100.00\\
    \hline
  \end{tabular}
  \caption{Accumulative \% of H-bonds whose predicted rotation values lie within the specified distance from the true rotation values, for the simple substitution matrices with different gap scores (see \cref{sec:method_align} for how the definition of these matrices). The numbers in parentheses show the volume of the ball whose radius is the upper limit of the range as a proportion of the volume of entire SO(3), w.r.t.\ the invariant metric.}
  \label{tab:align_gap}
\end{table*}

The results of the prediction using H-bond length sequence alignment with the substitution matrices $M^{(1)}$ to $M^{(4)}$ are shown in \Cref{tab:align_subst}. With the substitution matrix $M^{(1)}$, about half (49.56\%) of the predictions were placed inside a ball centred around the true rotation, occupying just 0.1\% of the volume of SO(3) (\Cref{fig:so3_a}), and 72.91\% inside a ball occupying 1\% of the SO(3) volume (\Cref{fig:so3_b}). We see that modifying the penalty for (``short to short'') substitution appears to have positive effect on the prediction accuracy compared to the linear penalty ($M^{(1)}$), and the improvements appear to be similar for the different penalties ($M^{(2)}$ to $M^{(4)}$). 

The effect of different gap scores using the simple substitution matrix $M_{\text{simple}}$ is shown in \Cref{tab:align_gap}. Perhaps surprisingly, the prediction accuracy using $M^{(-1)}_{\text{simple}}$ was very similar to the result obtained by using $M^{(4)}$, the substitution matrix with a large twistedness penalty (\Cref{tab:align_subst}). There was a reduction in accuracy when the gap score was set to 0, and a small improvement when it was set to -5.

The results from the pattern matching are shown in \Cref{tab:locpat_adj} under the column ``Orig. score''. We see in 61.10\% of all cases, the predicted rotation lies within a ball comprising just 0.1\% of the total volume of SO(3) centred at the true rotation, and in 86.90\% of all cases, the prediction was within a ball corresponding to 1\% of the volume of SO(3). To analyse the prediction results further, we looked at the DSSP classes \cite{kabsch83} of four residues around each H-bond. DSSP assigns seven secondary structure classes (plus ``unclassified'') to each residue (\Cref{tab:dssp}). 

\begin{table*}[t]
  \centering
  \begin{tabular}{*{6}{r}}
    \hline
    \multicolumn{2}{c}{Range of $d$} & Orig. score  & New score & Tertiary & 2020 data\\
    \hline
    $d < 0.2664$ & (0.1\%)  & 61.10 & 62.74 & 53.16 & 62.63 \\
    $d < 0.4567$ & (0.5\%)  & 80.72 & 82.17 & 74.51 & 82.42 \\
    $d < 0.5766$ & (1.0\%)  & 86.90 & 88.14 & 82.01 & 88.54 \\
    $d < 0.7862$ & (2.5\%)  & 92.87 & 93.80 & 89.70 & 93.97 \\
    $d < 0.9968$ & (5.0\%)  & 95.67 & 96.11 & 93.60 & 96.34 \\
    $d < 1.2689$ & (10.0\%) & 97.38 & 97.62 & 96.16 & 97.76 \\
    $d < 1.7663$ & (25.0\%) & 98.89 & 98.95 & 98.16 & 98.99 \\
    $d < 2.3099$ & (50.0\%) & 99.69 & 99.72 & 99.69 & 99.80 \\
    $d < 2.7437$ & (75.0\%) & 99.87 & 99.90 & 99.95 & 99.93\\
    $d < 3.1416$ & (100\%)  &100.00 & 100.00 & 100.00 & 100.00\\
    \hline
  \end{tabular}
  \caption{Accumulative \% of H-bonds whose predicted rotation values lie within the specified distance from the true rotation values, with the original score function \eqref{eq:score}, with the new score function \eqref{eq:score2}, with tertiary bonds, and training only with data from 2020 to predict newly added proteins.}
  \label{tab:locpat_adj}
\end{table*}

\begin{table*}
  \centering
  \begin{tabular}{*{3}{c}}
    \hline
    DSSP class & DSSP symbol & Local structure \\
    \hline
    $\alpha$-helix & H & \multirow{3}{*}{Helix} \\
    $3_{10}$-helix & G & \\
    $\pi$-helix & I & \\
    \hline
    Strand & E & Sheet \\
    \hline
    Isolated $\beta$-bridge residue & B & \multirow{3}{*}{Coil} \\
    Turn & T & \\
    Bend & S & \\
    \hline
    Unclassified & - & Unclassified \\
    \hline
  \end{tabular}
  \caption{DSSP classes and corresponding local structure patterns.}
  \label{tab:dssp}
\end{table*}

We analysed the frequencies of four-tuples of DSSP classes and the associated prediction accuracies (Supplementary Material), and observed the frequencies concentrated on a few classes, which is also evident if we look at the ten most frequent DSSP class combinations (\Cref{tab:4tuple}). We also observe that the residue class combination for the sheet structure (``EEEE'') has the high frequency and relatively low accuracy.

\begin{table}
  \centering
  \begin{tabular}{crr}
    \hline
    Residues & Frequency & Accuracy \\
    \hline
    HHHH & 9296 & 99.83\% \\
    EEEE & 5604 & 79.59\% \\
    HH-H & 780 & 99.36\% \\
    HTHH & 694 & 97.84\% \\
    T--T & 405 & 80.74\% \\
    TTHH & 390 & 96.15\% \\
    T-HH & 339 & 88.79\% \\
    GG-G & 290 & 93.45\% \\
    TS-T & 281 & 87.54\% \\
    H-HH & 263 & 91.25\% \\
    \hline
  \end{tabular}
  \caption[Most frequent DSSP combinations.]{10 most frequent combinations of DSSP classes around H-bonds, together with the proportions of predictions, which lie inside a ball centred at the true rotation having a volume corresponding to 1\% of the total volume of SO(3).}
  \label{tab:4tuple}
\end{table}

We then analysed each prediction and looked at the parameters used to produce it. By looking at the distance between predicted and true rotations ($\Delta$) and the mean distance to cluster mode ($m$) for each prediction, we found a group of predictions made using small window sizes, with many of them having large $\Delta$ values (Supplementary Material). This could happen, for example, if the H-bond pattern matched with a smaller window size has the associated cluster with a well-defined ``peak'' (thus having a low $m$ value and high score), while the cluster for a match with a larger window size has a lower ``peak'' (and a high $m$ value). The prediction was repeated with the modified score function (\ref{eq:score2}), with the results shown in \Cref{tab:locpat_adj}. We see the prediction accuracy was improved to 89.05\% inside 1\% SO(3) volume. Analysis of the other parameters (Remote bonds, Twisted bonds, and Residue groups) did not show any similar anomalies, and applying similar modifications to the score function to prioritise matches with more detailed patterns did not result in an improvement (results not shown).

\Cref{tab:locpat_adj} also shows the results from the analysis, where the H-bond local patterns were generated including the tertiary contact information. We see that 82.01\% of all predictions lay within 1\% SO(3)-volume of true values.

Furthermore, we ran the pattern matching prediction algorithm with the modified score function (\ref{eq:score2}) again, but only using PDB data from March 2020 for training, and and using 200 of the proteins added to PDB after March 2020 for validation. The results are shown in the column ``2020 data'' in \Cref{tab:locpat_adj}, and demonstrates the stability of our method with new structural data.

\section{Discussion}
In the first method, where H-bond pattern alignment score was used to find the best match, we were able to achieve over 70\% of the predictions within 1\% volume of SO(3) space from the true rotations. Changing the substitution matrices to simulate non-linear penalties did not result in any significant change in prediction accuracy. A minor improvement was observed when the penalty for the twistedness was increased (\Cref{tab:align_subst}). This may indicate the fact that the twistedness of an H-bond is directly related to its associated rotation value; if a bond is twisted and another is not twisted, it is unlikely that the two have similar rotation values. More surprisingly, an improvement similar to using large twistedness penalty was observed when using the ``simple'' substitution matrix where a match is given the score of 1 and a mismatch -1. A further, even larger improvement was observed by using the gap score of -5. As the window size is constant for the training and test datasets, larger gap score has the effect of making the algorithm more like the one that simply counts the match/mismatch between the two H-bond sequences. It is however not immediately clear why this may increase the prediction accuracy compared to allowing gaps, or making penalties dependent on the change in the bond lengths. It should be noted, that even though the Needleman-Wunsch algorithm itself does not place any restriction on its use, in the study of protein structures it is typically used for aligning segments of primary sequences. Accordingly, there are substitution matrices which are generally accepted in the field and tested in various applications \cite{mount08}. The sequences considered in this study have not been studied previously, and the substitution matrix was developed specifically for this study. It is therefore possible that a further development of the substitution matrix may improve the prediction accuracy. 

In the second method, where the rotation prediction was done by finding an exact match for a given H-bond pattern, we were able to achieve close to 90\% of our predictions lying inside 1\% SO(3) volume of the true rotations. We believe there are potentials for further improvement, since an analysis of the clustering results show that if we could choose the ``best'' clustering result, i.e.\ the clustering result that lies closest to the true value, it will result in over 95\% and 80\% of ``predictions'' inside 1\% and 0.1\% SO(3)-volume, respectively. So it is possible that a better score function than \eqref{eq:score2} may give the necessary improvement. An improvement may also be possible by a method based on a machine-learning approach, which has proved so successful in protein structure prediction. It is also clear from the results of the DSSP class analysis, that we need to improve our predictions in the sheet structure, if we are to reach our goal. Our hypothesis is that due to the structural flexibility of sheets compared to helices, the bonds in sheets are affected by the nearby atoms (which are not necessarily near the bond in the secondary structure) to a greater extent than the bonds in helices. We have attempted, unsuccessfully so far, to take this into account by using tertiary bond information in our H-bond pattern generation. A better understanding of $\beta$-sheet structures and their topology may be needed to improve prediction of H-bond rotations inside $\beta$-sheets.

\subsubsection*{Acknowledgement}
This paper is partly a result of the ERC-SyG project, Recursive and Exact New Quantum Theory (ReNewQuantum) which received funding from the European Research Council (ERC) under the European Union's Horizon 2020 research and innovation programme under grant agreement No. 810573.

\printbibliography

\end{document}


\beginsupplement
\section{Supplementary Material}

\subsection{Spatial rotation along a H-bond}

\begin{figure*}[h]
  \centering
  \includegraphics[scale=.5]{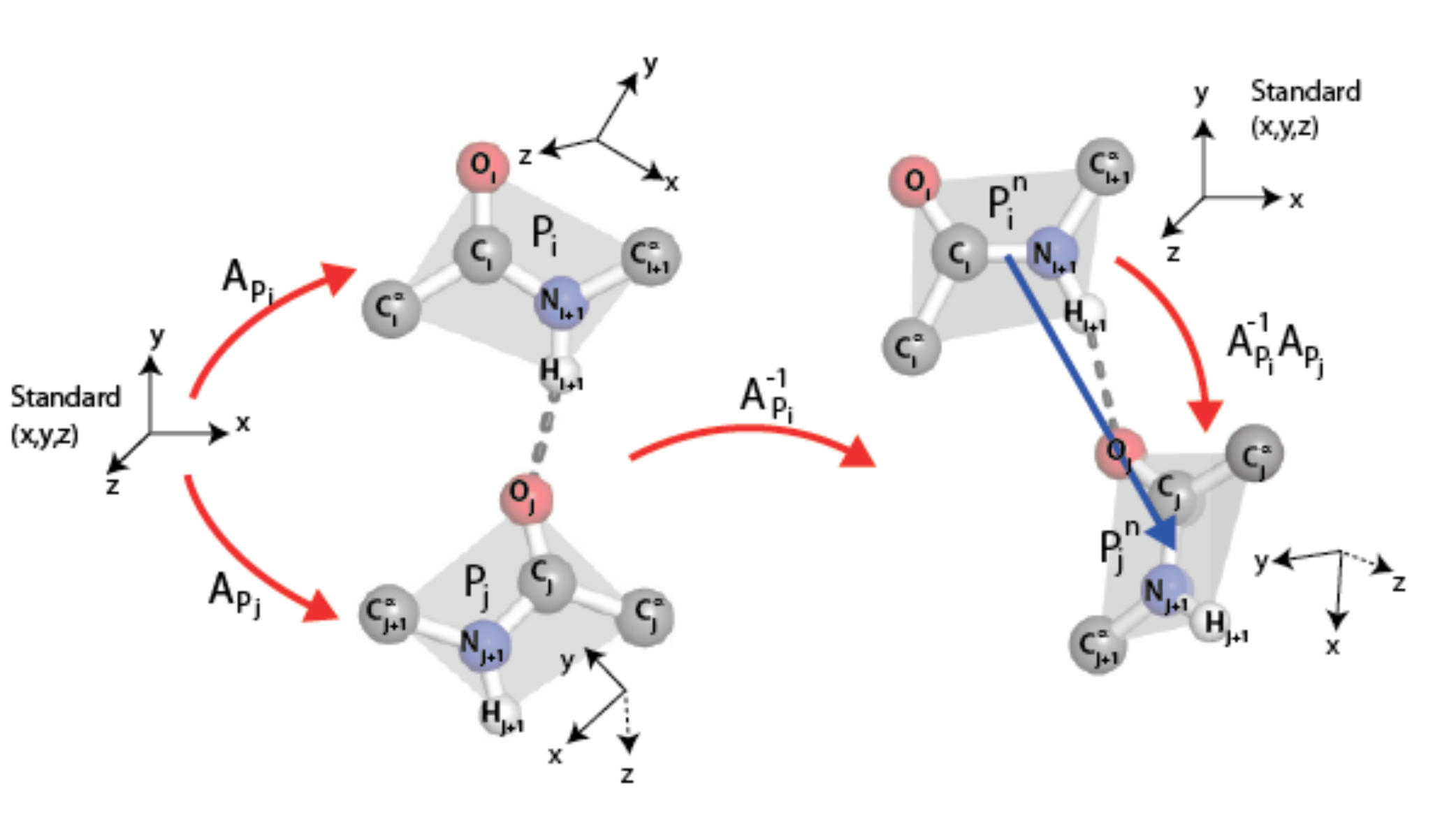}
  \caption{Assignment of SO(3) element to a H-bond. See text for details.}
  \label{fig:so3}
\end{figure*}

We give a short summary of the assignment of SO(3) element to a H-bond, as illustrated in \Cref{fig:so3}. First we associate a triple of orthogonal unit vectors $\mathcal{F}_i = (u_i, v_i, w_i), \;u_i,v_i,w_i \in \R^3$ to the $i$'th peptide unit $P_i$, $i=1,2,\dotsc$ as follows. $u_i$ is the unit vector parallel to the displacement vector  $\overrightarrow{\mathrm{C}_{i}\mathrm{N}_{i+1}}$ along the peptide bond. Let $u_i^\perp$ be the vector in the plane of peptide unit which is perpendicular to $u_i$. Let $\bar{v}_i$ be the projection of the vector $\overrightarrow{\mathrm{C}^\alpha_{i}\mathrm{C}_i}$ onto the subspace $\R u_i^\perp$ and set $v_i = \bar{v}_i/\norm{\bar{v}_i}$. Finally, we set $w_i = u_i \times v_i$, where $\times$ denotes the cross product. For each $i$, there is a unique element of SO(3) $A_{P_i}$, which takes the standard three frame $(x,y,z)$ to $\mathcal{F}_i$. Let $\mathcal{F}_j$ be the three frame associated to another peptide unit $P_j$. Then there is a unique element of the group $\mathrm{SO}(3)$, $A_{P_j}(A_{P_i})^{-1}$, which takes $\mathcal{F}_i$ to $\mathcal{F}_j$. However, this element of $\mathrm{SO}(3)$ is not invariant under rotation of the entire protein. The solution to this problem is to apply the rotation $A_{P_i}^{-1}$ to $\mathcal{F}_i$ and $\mathcal{F}_j$, so that $\mathcal{F}_i$ becomes the standard three-frame in $\R^3$. The element of SO(3) that takes the standard three-frame to $A_{P_i}^{-1}\mathcal{F}_j$ is $A_{P_i}^{-1}A_{P_j}$. If there is a H-bond from $P_i$ to $P_j$ in our model, we assign $A_{P_i}^{-1}A_{P_j} \in \mathrm{SO}(3)$ to this edge.

\pagebreak

\subsection{Grouping of amino-acid residues for construction of H-bond local patterns}
When constructing H-bond local patterns, we divide the 20 ``standard gene-encoded'' amino acids into up to four groups according to the scheme shown in \Cref{tab:amino_groups}.

\renewcommand{\arraystretch}{1.5}
\begin{table}[h]
  \setlength{\tabcolsep}{2pt}
  \centering
  \begin{tabular}[c]{l|c|*{21}{l}}
    \multicolumn{2}{c|}{\rotatebox[origin=l]{90}{Amino acids}}
    & \rotatebox[origin=l]{90}{(L) Leucine}
    & \rotatebox[origin=l]{90}{(V) Valine}
    & \rotatebox[origin=l]{90}{(I) Isoleucine}
    & \rotatebox[origin=l]{90}{(F) Phenylalanine}
    & \rotatebox[origin=l]{90}{(M) Methionine}      
    & \rotatebox[origin=l]{90}{(A) Alanine}
    & \rotatebox[origin=l]{90}{(G) Glycine}
    & \rotatebox[origin=l]{90}{(S) Serine}
    & \rotatebox[origin=l]{90}{(C) Cysteine}
    & \rotatebox[origin=l]{90}{(E) Glutamic acid}
    & \rotatebox[origin=l]{90}{(K) Lysine}
    & \rotatebox[origin=l]{90}{(R) Arginine}
    & \rotatebox[origin=l]{90}{(D) Aspartic acid}
    & \rotatebox[origin=l]{90}{(T) Threonine}
    & \rotatebox[origin=l]{90}{(Y) Tyrosine}
    & \rotatebox[origin=l]{90}{(N) Asparagine}
    & \rotatebox[origin=l]{90}{(Q) Glutamine}
    & \rotatebox[origin=l]{90}{(H) Histidine}
    & \rotatebox[origin=l]{90}{(W) Tryptophan}
    & \rotatebox[origin=l]{90}{(P) Proline}
    & \rotatebox[origin=l]{90}{(X) N/A} \\
    \hline
    \multirow{4}{*}{\rotatebox[origin=l]{90}{No.of groups}}
    & 1 & \multicolumn{21}{c}{X}\\
    \cline{2-23}
    & 2 & \multicolumn{9}{c|}{L} & \multicolumn{11}{c|}{E} & X \\
    \cline{2-23}
    & 3 & \multicolumn{5}{c|}{L} & \multicolumn{4}{c|}{A} & \multicolumn{11}{c|}{E} & X \\
    \cline{2-23}
    & 4 & \multicolumn{5}{c|}{L} & \multicolumn{4}{c|}{A} & \multicolumn{10}{c|}{E} & \multicolumn{1}{c|}{P} & X \\
  \end{tabular}
  \caption[Residue groups]{Grouping scheme for amino acids and group labels.}
  \label{tab:amino_groups}
\end{table}

\subsection{Analysis of prediction accuracy by the four residues around the central bond}
\begin{figure}[H]
  \centering
  \begin{subfigure}{\textwidth}
    \centering
    \includegraphics[scale=.9]{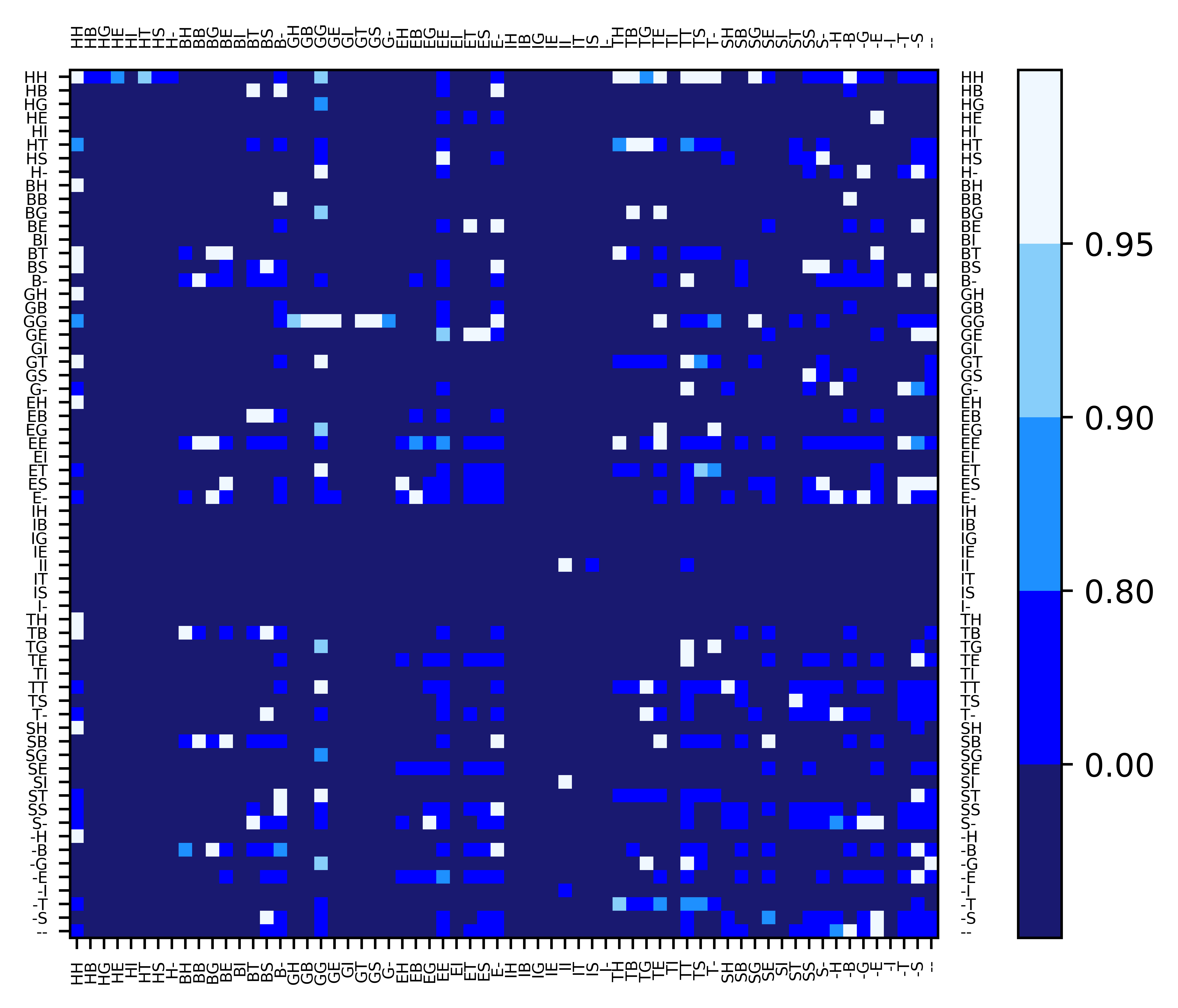}
    \caption{Accuracy of predictions by the DSSP classes of four residues around the H-bond; \% of predictions lying within 1\% SO(3) volume.}
    \label{fig:dssp_acc}
  \end{subfigure}
  \begin{subfigure}{\textwidth}
    \centering
    \includegraphics[scale=.9]{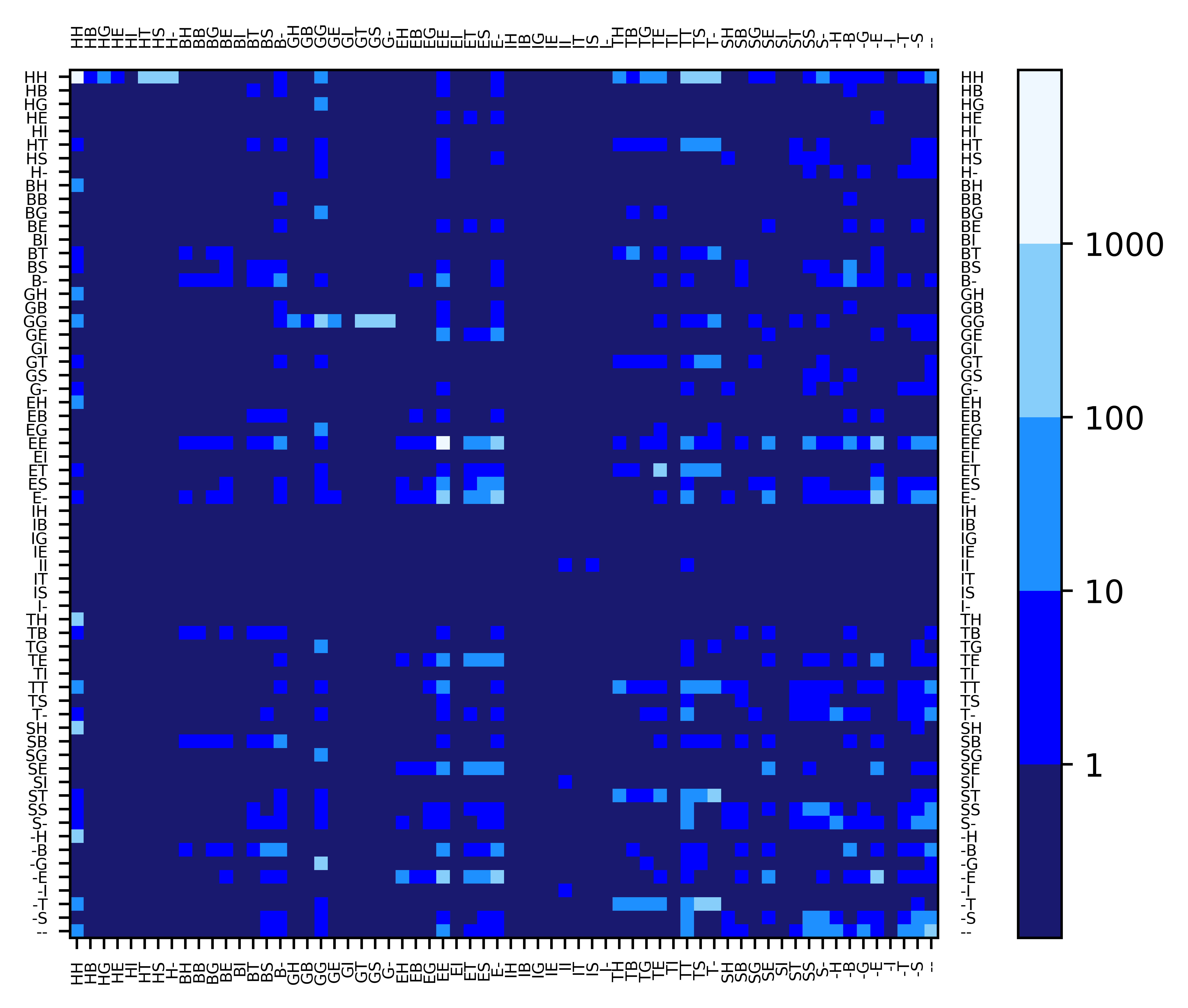}
    \caption{Frequency by the DSSP classes of four residues around the H-bond.}
    \label{fig:dssp_freq}
  \end{subfigure}
  \caption{Accuracy and frequency of predictions by DSSP classes. The horizontal and vertical axes show the donor-side and the acceptor-side residues, respectively.}
  \label{fig:dssp}
\end{figure}

We analysed accuracy (\Cref{fig:dssp_acc}) and frequency (\Cref{fig:dssp_freq}) of our predictions by the DSSP classes of four residues around the H-bonds. The residues are ordered along the backbone, so the two-letter combinations along the horizontal (donor-side) and the vertical (acceptor-side) axes show the DSSP class of the residue immediately preceding, followed by the residue immediately after, the H-bond donor (acceptor).

\pagebreak
\subsection{Relation between the prediction accuracy and the size of clusters}
\subsubsection{Analysis by window size}
\begin{figure}[h]
  \centering
  \includegraphics[width=\linewidth]{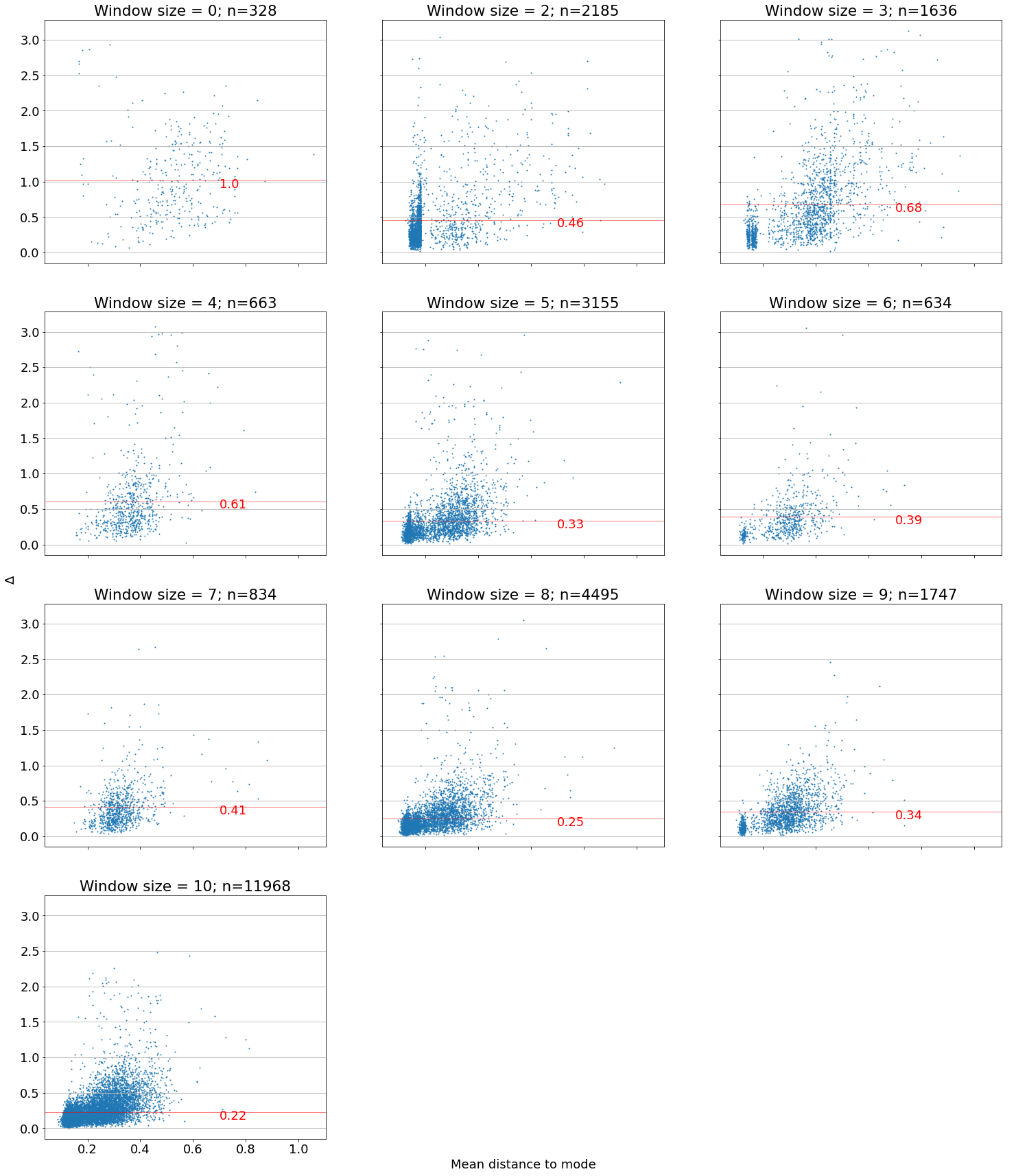}
  \caption[Delta vs m 2]{Distance to the true rotation ($\Delta$) and mean distance to cluster mode ($m$), filtered by the widnow size used for prediction. No prediction was made using window size 1, so the plot was omitted. The red line in dicates the mean for each plot.}
  \label{fig:scatter_win1}
\end{figure}

\begin{figure}[h]
  \centering
  \includegraphics[width=\linewidth]{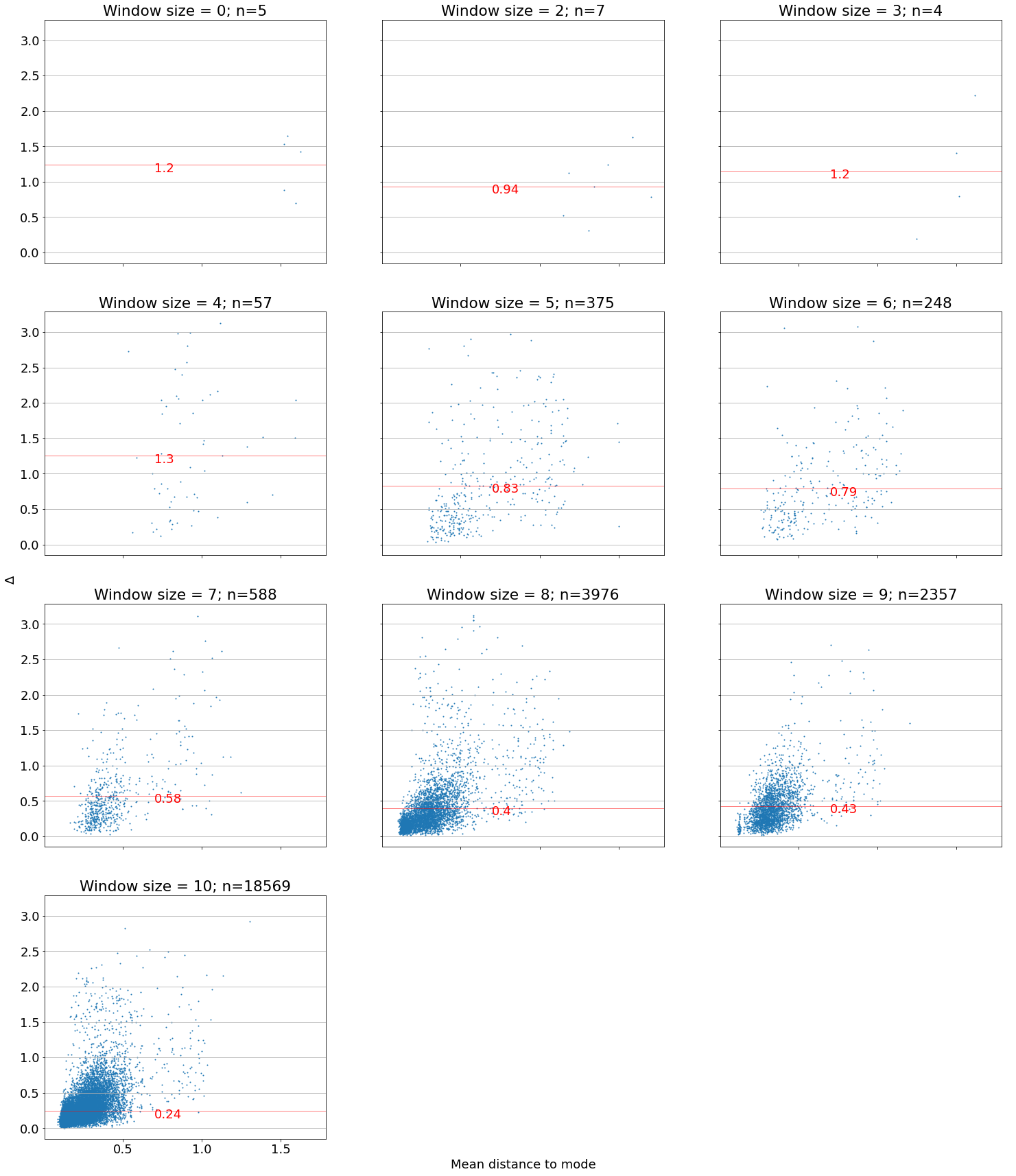}
  \caption[Delta vs m 3]{Distance to the true rotation ($\Delta$) and mean distance to cluster mode ($m$), filtered by the widnow size used for prediction. No prediction was made using window size 1, so the plot was omitted. The data is obtained by the modified score function (3). The red line in dicates the mean for each plot.}
  \label{fig:scatter_win2}
\end{figure}

We analysed the relation between the cluster scores, distance between the predicted and true rotations, and the window size used for prediction, using the results obtained by the original score function (2). \Cref{fig:scatter_win1} shows the distribution of the scores (here using the mean distance to the mode box, or $m$ in (2)) against the distance between the predicted and true rotations ($\Delta$), filtered by the window size used for prediction. A cluster of predictions with low $m$ values and high $\Delta$ values can be seen in window size 2, and to a smaller extent in window size 3. We modified the score function to (3), to promote the selection of larger pattern for prediction. This had the desired effect on the prediction results, with few predictions made by small patterns (\Cref{fig:scatter_win2}).

\subsubsection{Analysis by remote bond}
\begin{figure}[h]
  \centering
  \includegraphics[width=\linewidth]{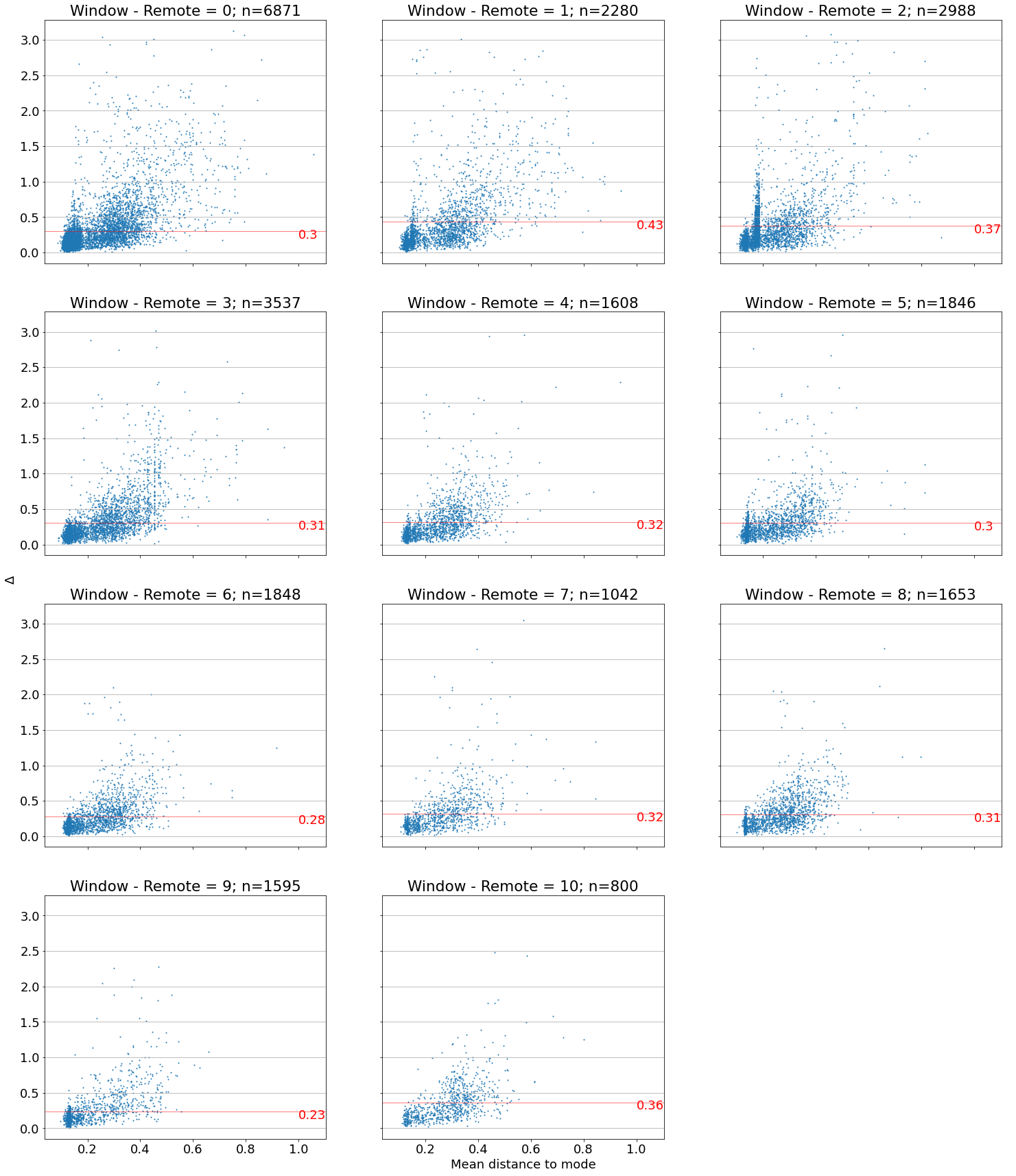}
  \caption[Delta vs m 2]{Distance to the true rotation ($\Delta$) and mean distance to cluster mode ($m$), filtered by the difference between the widnow size and remote bond parameter used for prediction. The red line in dicates the mean for each plot.}
  \label{fig:scatter_rem1}
\end{figure}

\begin{figure}[h]
  \centering
  \includegraphics[width=\linewidth]{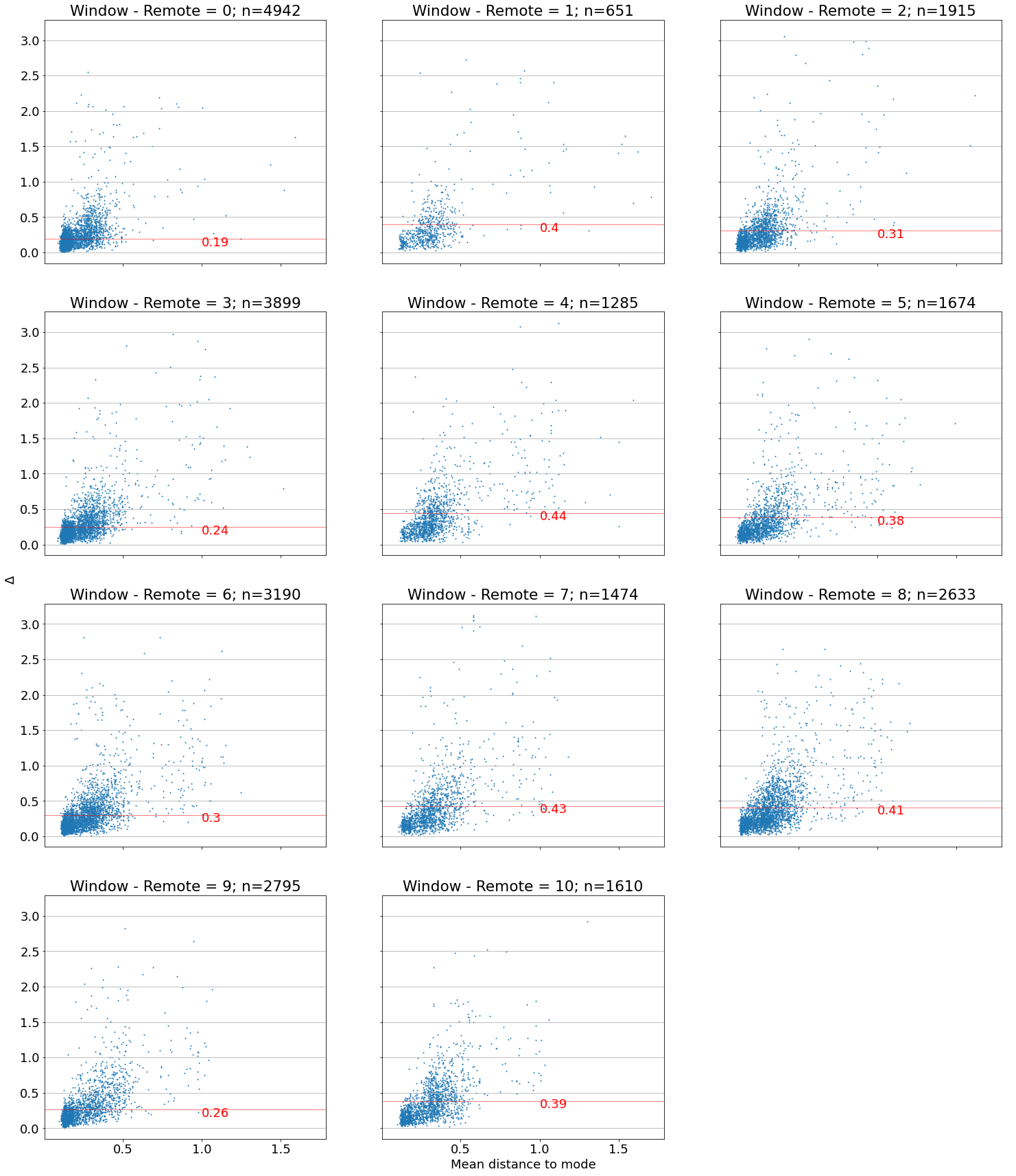}
  \caption[Delta vs m 3]{Distance to the true rotation ($\Delta$) and mean distance to cluster mode ($m$), filtered by the difference between the widnow size and remote bond parameter used for prediction. The data is obtained by the modified score function (3). The red line in dicates the mean for each plot.}
  \label{fig:scatter_rem2}
\end{figure}

We analysed the relation between the cluster scores, distance between the predicted and the true rotations, and the remote bond parameter used for predictions. \Cref{fig:scatter_rem1} shows the distribution of $m$ against $\Delta$, filtered by the difference between window size and remote bond parameters used for prediction, obtained by the score function (2). Similar clusters with low $m$ and high $\Delta$ values can be observed for the difference of 2. The cluster is linked to the one observed in \Cref{fig:scatter_win}, where the remote bond parameter is 0. This can be verified by running the same analysis, but this time with the results obtained by the modified score function (3). Indeed, such cluster can not be observed with the data from the score function (3) (\Cref{fig:scatter_rem2}.

\subsubsection{Analysis by twistedness}
\begin{figure}[h]
  \centering
  \includegraphics[width=\linewidth]{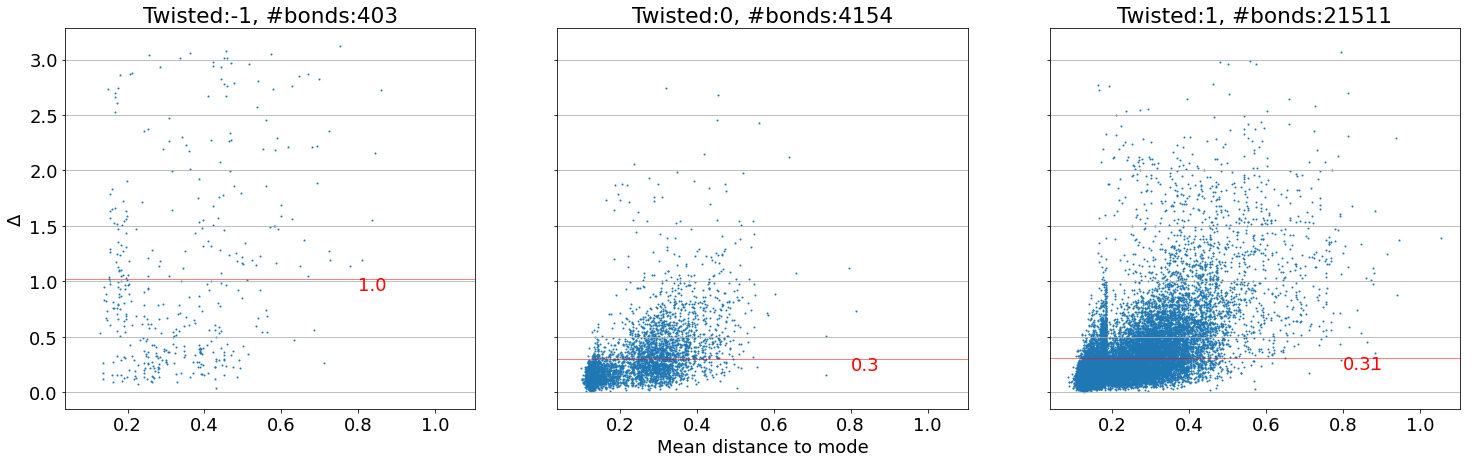}
  \caption[Delta vs m 2]{Distance to the true rotation ($\Delta$) and mean distance to cluster mode ($m$), filtered by the twistedness parameter used for prediction (-1: no twisted bond recorded, 0: twistedness of central bond only, 1: twistedness of all bonds). The red line in dicates the mean for each plot.}
  \label{fig:scatter_twi1}
\end{figure}

\begin{figure}[h]
  \centering
  \includegraphics[width=\linewidth]{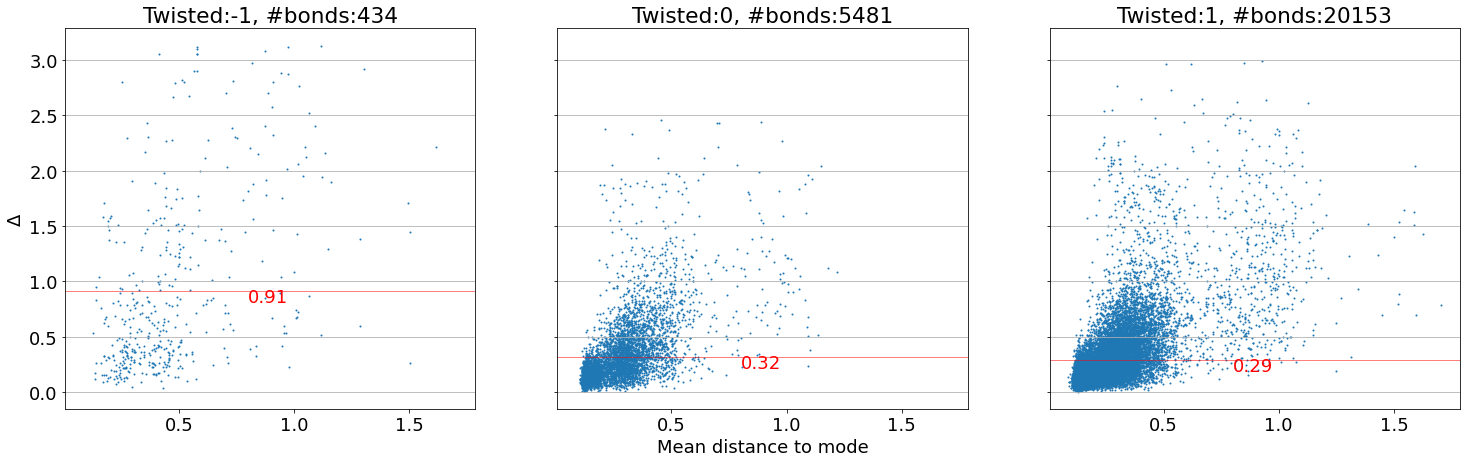}
  \caption[Delta vs m 3]{Distance to the true rotation ($\Delta$) and mean distance to cluster mode ($m$), filtered by the twistedness parameter used for prediction (-1: no twisted bond recorded, 0: twistedness of central bond only, 1: twistedness of all bonds). The data is obtained by the modified score function (3). The red line in dicates the mean for each plot.}
  \label{fig:scatter_twi2}
\end{figure}

We analysed the relation between the cluster scores, distance between the predicted and the true rotations, and the twistedness parameter used for predictions. We can again observe a cluster with low $m$ and high $\Delta$ values in the data produced with the original score function ((2); \Cref{fig:scatter_twi1}), which is no longer present when the modified score function (3) is used (\Cref{fig:scatter_twi2}. 

\subsubsection{Analysis by residue groups}
\begin{figure}[h]
  \centering
  \includegraphics[width=\linewidth]{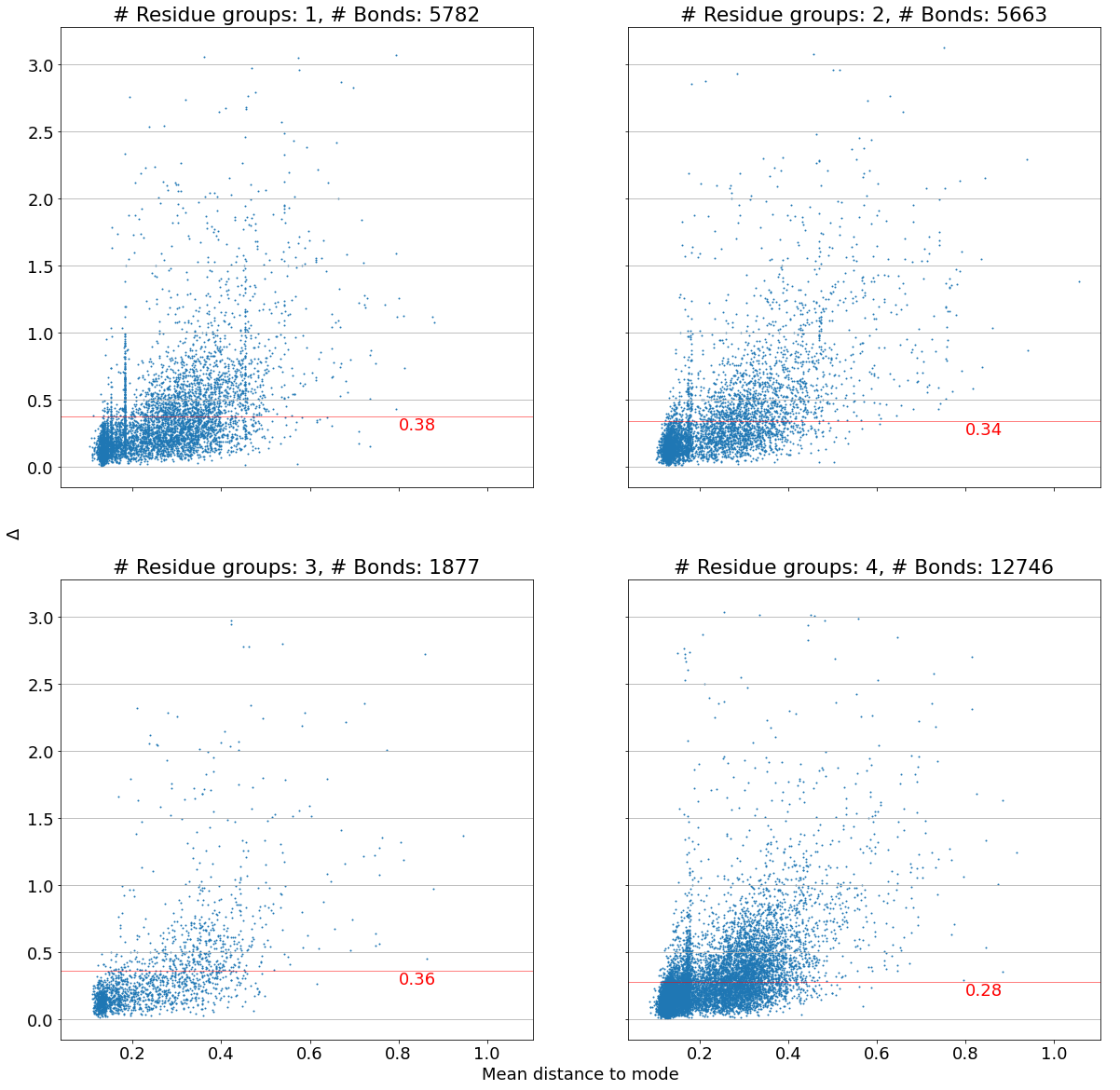}
  \caption[Delta vs m 2]{Distance to the true rotation ($\Delta$) and mean distance to cluster mode ($m$), filtered by the residue group parameter used for prediction (the parameter stores the number of groups used to classify residues). The red line in dicates the mean for each plot.}
  \label{fig:scatter_res1}
\end{figure}

\begin{figure}[h]
  \centering
  \includegraphics[width=\linewidth]{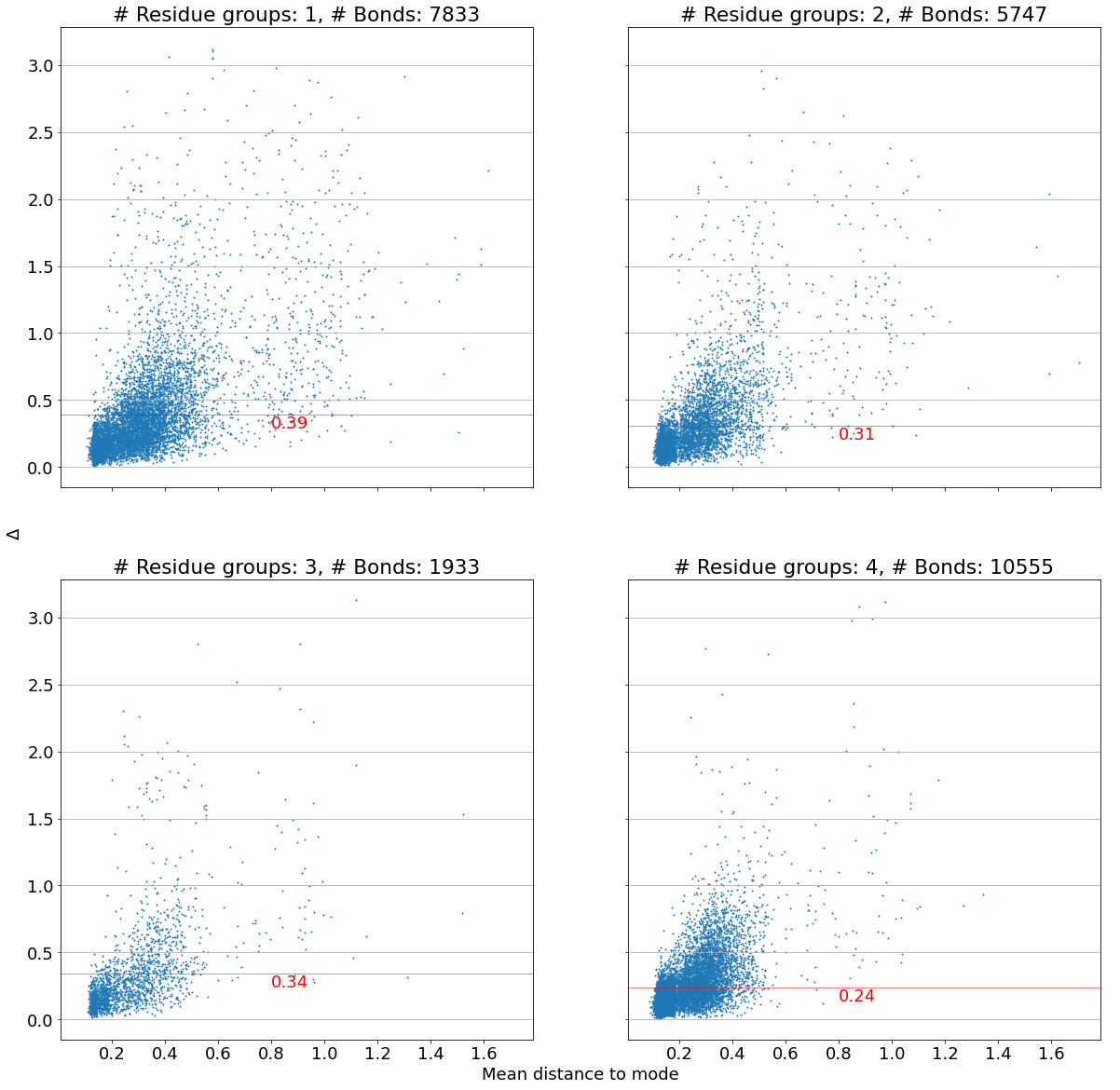}
  \caption[Delta vs m 3]{Distance to the true rotation ($\Delta$) and mean distance to cluster mode ($m$), filtered by the residue group parameter used for prediction (the parameter stores the number of groups used to classify residues). The data is obtained by the modified score function (3). The red line in dicates the mean for each plot.}
  \label{fig:scatter_res2}
\end{figure}

We analysed the relation between the cluster scores, distance between the predicted and the true rotations, and the residue group parameter used for predictions. We can again observe a cluster with low $m$ and high $\Delta$ values in the data produced with the original score function ((2); \Cref{fig:scatter_res1}), which is no longer present when the modified score function (3) is used (\Cref{fig:scatter_res2}. 


\clearpage
\printbibliography